\documentclass[%
 aip, amsmath,amssymb,
 preprint,%
 ]{revtex4-1}

\usepackage{graphicx}
\usepackage{dcolumn}% Align table columns on decimal point
\usepackage{bm}% bold math
\usepackage[mathlines]{lineno}% Enable numbering of text and display math
\usepackage{booktabs}

\usepackage{siunitx}
\usepackage{epstopdf}
\usepackage{color}
\usepackage[sort&compress]{natbib}
\usepackage[capitalise]{cleveref}

\begin{document}

\preprint{AIP/123-QED}

\title{Thermomechanical response of thickly tamped targets and diamond anvil cells under pulsed hard x-ray irradiation
%[Thermal response/Differential heating] [and thermal evolution] of [highly/strongly] tamped [multi-layer] [targets/samples/foils] under high-brightness [hard x-ray] [pulsed] irradiation [by [bright/high-brightness/[tight-]focussed] hard x-rays].
}

\author{J. Meza-Galvez}
\affiliation{Facultad de Qu\'imica, Universidad Aut\'onoma del Estado de M\'exico (UAEM\'ex), Tollocan s/n, esq. Paseo Col\'on, Toluca, Estado de M\'exico, 50110, M\'exico.}
\affiliation{The School of Physics and Astronomy, Centre for Science at Extreme Conditions, and SUPA, University of Edinburgh, Peter Guthrie Tait Road, Edinburgh, EH9 3FD, UK}

\author{N. Gomez-Perez}
%\email{ngomez@uniandes.edu.co}
%\homepage{http://www2.ph.ed.ac.uk/~v1ngomez/}
\affiliation{The School of Physics and Astronomy, Centre for Science at Extreme Conditions, and SUPA, University of Edinburgh, Peter Guthrie Tait Road, Edinburgh, EH9 3FD, UK}
\affiliation{School of Mathematics, Statistics and Physics, Newcastle University, Newcastle upon Tyne, NE1 7RU, UK}
\affiliation{British Geological Survey, Currie, EH14 4BA, UK}

\author{A. Marshall}
\affiliation{%I
The School of Physics and Astronomy, Centre for Science at Extreme Conditions, and SUPA, University of Edinburgh, Peter Guthrie Tait Road, Edinburgh, EH9 3FD, UK}

\author{A.L. Coleman}
\affiliation{%
The School of Physics and Astronomy, Centre for Science at Extreme Conditions, and SUPA, University of Edinburgh, Peter Guthrie Tait Road, Edinburgh, EH9 3FD, UK}

\author{K. Appel}
\affiliation{European XFEL GmbH, Holzkoppel 4, 22869 Schenefeld, Germany}

\author{H.P. Liermann}
\affiliation{Deutsches Elektronen-Synchrotron (DESY) Photon Science, Notkestra$\ss$e 85, 22607 Hamburg, Germany}

\author{M. I. McMahon}
\affiliation{%
The School of Physics and Astronomy, Centre for Science at Extreme Conditions, and SUPA, University of Edinburgh, Peter Guthrie Tait Road, Edinburgh, EH9 3FD, UK}

\author{Z. Kon\^opkov\'a}
\affiliation{European XFEL GmbH, Holzkoppel 4, 22869 Schenefeld, Germany}
%European XFEL GmbH, Notkestra$\ss$e 85, DE-22607 Hamburg, Germany}

%
\author{R. S. McWilliams}
\email{rs.mcwilliams@ed.ac.uk}
%\homepage{http://www2.ph.ed.ac.uk/~rmcwilli/}
\affiliation{%
The School of Physics and Astronomy, Centre for Science at Extreme Conditions, and SUPA, University of Edinburgh, Peter Guthrie Tait Road, Edinburgh, EH9 3FD, UK}

\date{\today}

\begin{abstract}

In the laboratory study of extreme conditions of temperature and density, the exposure of matter to high intensity radiation sources has been of central importance.  Here we interrogate the performance of multi-layered targets in experiments involving high intensity, hard x-ray irradiation, motivated by the advent of extremely high brightness hard x-ray sources, such as free electron lasers and 4$^{\text{th}}$-generation synchrotron facilities.  Intense hard x-ray beams can deliver significant energy in targets having thick x-ray transparent layers (tampers) around samples of interest, for the study of novel states of matter and materials' dynamics.  Heated-state lifetimes in such targets can approach the microsecond level, regardless of radiation pulse duration, enabling the exploration of conditions of local thermal and thermodynamic equilibrium at extreme temperature in solid density matter.  The thermal and mechanical response of such thick layered targets following %differential 
x-ray heating, 
%from energy delivery and local lattice equilibration on femto- to picosecond timescales 
including hydrodynamic relaxation and heat flow on picosecond to millisecond timescales, is modeled using radiation hydrocode simulation, finite element analysis, and thermodynamic calculations.
%Timescales and lengthscales are much larger than those of typical electron equilibration processes, and equilibrium is thus a reasonable first approximation for models of relevant phenomena including heat transport and hydrodynamic effects.
Assessing the potential for target survival over one or more exposures, and resistance to damage arising from heating and resulting mechanical stresses, this study doubles as an investigation into the performance of diamond-anvil high pressure cells under high x-ray fluences.  Long used in conjunction with synchrotron  x-ray radiation and high power optical lasers, the strong confinement afforded by such cells suggests novel applications at emerging high intensity x-ray facilities 
%free electron lasers and modern synchrotron beamlines, 
and new routes to studying thermodynamic equilibrium states of warm, very dense matter.% via the capacity to confine extreme states for particularly long durations
%Indeed, we find that configuring a tamped target as a high pressure cell confers advantages including control of sample expansion, resistance thermal stresses and efficient control of heat.
%This study suggests .
%equilibration processes to proceed, from approach to local thermal equilibrium between electrons and ions to phase transformations to stable ionic structures.

%with thermal damage more likely when using higher Z materials as samples, low thermal conductivity tampers, and no intermediate layer between sample and tamper to act as a thermal barrier.

%We apply this analysis to the specific case of the diamond anvil high pressure cell and discuss how 
 %at the same time, such stresses can be controlled by adopting an anvil-cell type configuration where stresses are resisted and samples confined.

\end{abstract}

%PACS
%\verb+
%\pacs{#1}
%+ command
%\dots %Valid PACS numbers may be entered using the \verb+\pacs{#1}+ command.


\maketitle

\crefname{figure}{Fig.}{Figs.}
\crefname{equation}{Eq.}{Eqs.}

\section{\label{sec:intro} INTRODUCTION}

% \SI{1000}{\kelvin} 

Matter with an atomic density similar to that of the solid state, at temperatures of thousands to millions of degrees Kelvin and pressures exceeding millions of atmospheres, and undergoing rapid changes on microsecond to femtosecond timescales, %on fast (microsecond-nanosecond) to ultrafast (picosecond to femtosecond) timescales, 
is central to our understanding of planetary and stellar interiors, fusion energy technologies, and fundamental materials physics and chemistry.  
%Such ?high energy density? conditions include states of 
These warm dense matter states are not well described by the theoretical simplifications of traditional condensed matter physics or plasma physics. %, that common theoretical simplifications are often not applicable. 
Laboratory experiments are thus critical for developing a physical understanding of this regime of temperature, density, pressure, and timescale.
The creation and probing of warm dense matter in the laboratory often relies on central facilities capable of delivering high-brilliance irradiation, which can rapidly generate extreme temperatures in dense (i.e. solid or liquid) targets by ultrafast (fs-ps) `isochoric' heating, or by the production of dynamic compression waves within the target facilitated by the expansion of heated matter on longer (ps-ns) timescales\cite{ng:2005, Sentoku:2007, Fletcher:2014}.
 Ultrafast techniques have been widely employed to study the case of isochoric heating at timescales from femtosecond energy delivery to electrons, to picosecond heating of the lattice ions and hydrodynamic expansion into a vapor\cite{Patel:2003, Gregori:2005,Ivanov:2003,Levy:2015,Sentoku:2007}.
%USEDhttps://pdfs.semanticscholar.org/65cf/fbf6c0f7c9aad2c7f556b234a4ce2c14a02f.pdf?
%USED http://onlinelibrary.wiley.com/doi/10.1002/ctpp.200510032/abstract?

A common strategy uses electromagnetic radiation, often in the optical or UV range, to deliver the intense energy burst.  In such photonic experiments energy is delivered directly to electrons,
%optical rad papers, Sentoku:2007, Vinko2012 for x-ray, Nagler for x-ray
which then transfer energy to the ions (lattice) as the system relaxes toward a state of local thermal equilibrium (LTE), a prerequisite for reaching local thermodynamic equilibrium conditions.  The timescale of equilibration between the ions and electrons is typically %in the range of ps to ns
on the order order of ps\cite{White:2014,Zastrau:2014,Mostovych:1997,Ivanov:2003}.
%[
%USED ZastrauPRE14, 
%USEDWhitePRl2014
%ARCHAIC,SKIP: CelliersPRL92,
%USED MostovychPRL97,? 
%SKIP, GRAMMATICALSHIT: https://www.ncbi.nlm.nih.gov/pmc/articles/PMC4702138/#b24? 
%]; 
%Sentoku:2007
As electron-ion equilibration occurs roughly coincident with the expansion, melting, and vaporization processes naturally coupled to lattice heating, a loss of high-density conditions and sample confinement can occur before LTE is achieved, leading to study of nonequilibrium matter exclusively. % and the ionic and electronic structures adjust to an equilibrium state. %(typically, but not necessarily, those of a disordered fluid).  
The experimental timescale is also controlled by the size of targets, which in high power but low photonic-energy experiments is limited by short radiation absorption lengths, even in dielectrics.  Such practical challenges of using radiation heating to study equilibrium warm dense matter in the laboratory often complicate the experimental study of equilibrium extreme systems common in nature and technology.  
Other methods of irradiative volumetric energy deposition providing access to similar states of matter have similar limitations, include intense proton\cite{Patel:2003,Ping:2015}, heavy ion\cite{Tahir:2017},
%SKIP: Ni et al. Laser and Particle Beams 2008, 
%USED Tahir.Ion.Beam.Extremes.ApJ.2016.pdf, http://iopscience.iop.org/article/10.3847/1538-4365/aa813e/pdf
%
and electron\cite{White:2014}
%USED  White PRL 2014
 beams. %or other modes of fast electron deposition
% , e.g. from short pulse laser interactions 
% \cite{Gregori:2005, Ping:2006, Sentoku:2007}
%NO, its mostly a shock paper: http://iopscience.iop.org/article/10.1209/0295-5075/119/35001/meta
%and dynamic compression waves 
%\cite{
%Falk:2018}.
%USED:https://journals.aps.org/prl/pdf/10.1103/PhysRevLett.120.025002
%Ehhhh,NO. \& other paper before?,
Dynamic compression,
%itself,
the driving of compression (i.e. shock or ramp) waves traveling at near sound velocities ($\sim$1-10 $\mu$m/ns) \cite{ng:2005, McWilliams:2010,Fletcher:2014,Kraus:2017, Falk:2018, Eggert:2010}, is a somewhat slower form of volumetric energy delivery,
%heating (and compression),
while diffusive\cite{Goncharov:2010, Goncharov:2012, McWilliams:2015, Konopkova:2016}  (as opposed to ballistic\cite{Gregori:2005, Ping:2006, Sentoku:2007}) heat conduction %through targets 
is even slower.  While these latter approaches in principle provide better access to equilibrium states of warm dense matter, they are limited by restriction to adiabatic pathways (dynamic compression) and by the aforementioned challenges of confining very hot matter (diffusion).

One strategy to extend the lifetime of an irradiation-driven warm dense state is to provide a tamper material around samples through which energy may be deposited and which delay, prevent, or otherwise control expansion\cite{Tahir:2017,Mostovych:1997,Ping:2015,Vailionis:2011,Johnson:2005,Bailey:2003},
%USEDMostovychPRL97, PingPoP15, VailionisNatComm11, 
%NO https://www.nature.com/articles/ncomms8555
%NO? https://journals.aps.org/prl/abstract/10.1103/PhysRevLett.82.4843?, 
%YES https://journals.aps.org/prl/pdf/10.1103/PhysRevLett.94.057407, 
%YES http://www.sciencedirect.com/science/article/pii/S0022407303000505 (GOOD), 
%JUST1: IONBEAM_CYLINDER1/2
 such as by extending the time it takes pressure release waves and cracks to propagate through the heated target.
%  , which mediate the expansion of the target, 
%Expansion produces, at sufficiently high temperatures, a gas or diffuse plasma.  
This tamping approach can even confine the heated region entirely, enabling recovery of high density samples quenched from conditions that would normally lead to vaporization \cite{Vailionis:2011}. %[
%VailionisNatComm11+ below
%] 
%Rand allow further experiments on the same target.
For optical radiation, tamping can be achieved by placing an absorptive (i.e. metal) layer between transparent (i.e. dielectric) tamper materials\cite{Mostovych:1997,Ping:2015,Johnson:2005},
%USED Ping15, MostovychPRL97, transpose relevant from above list - JOhnson so far...
by tightly-focussing the beam within the tamper itself\cite{Vailionis:2011}, % [
%USED VailionisNatComm11, 
%NO Gamaly, http://www.sciencedirect.com/science/article/pii/S1574181811000966
 or other configurations such as utilizing energetic electron transport to deposit energy deeply in a target\cite{Sentoku:2007}. However, tamping using high-power optical laser irradiation is limited by the need to deliver sufficient energy through the tamper to the sample, and thus depends on the optical transmission of the material under high brightness radiation, often requiring thin tampers %[NEED EXAMPLE FROM HT ISOCHORIC study] 
at all but the lowest
irradiances\cite{Mostovych:1997} which limit the efficacy of this strategy.
 %frequently producing similarly short-lived states due to a combination of the thin tamper configuration and due to fundamental limitations on optical radiation absorption lengths, such as in natural absorbers such as metals and in insulators at high irradiance (i.e. dielectric breakdown), which allow any thermomechanical exitation to rapidly dissipate. 
%Even under ideal conditions, the thickness of the excited region is typically limited to $\sim$tens of nm, with tampers up to $\sim$$\mu$m thickness  possible.  % spot sizes are diffraction-limited to several $\mu$m diameter or larger.  Such strategies typically ensure nonequilibrium conditions (i.e. of electrons and ion temperature) or prohibit thermodynamic equilibrium [studies/from being studied].
Targets of $\mu$m level thicknesses with experimental lifetimes of ps, set by unconfined hydrodynamic expansion, remain common.

Intense x-rays also rapidly heat matter\cite{Zastrau:2008,Nagler:2009,Vinko:2012,Ping:2015,Zastrau:2014,Fletcher:2014,Glenzer:2016,Levy:2015,Saunders:2018}. %,and this work focusses on the use of intense x-rays to heat samples
This energy deposition may be introduced deliberately (e.g. to heat or otherwise excite electrons in a sample% or to filter beam energy
) or may be a side effect of probing samples with a high intensity x-ray beam.  X-ray heating does not depend on damage thresholds of targeted materials, as in optical laser experiments, but instead depends nearly linearly on their x-ray absorption properties, %, with x-ray absorptivity
which depend %strongly 
 on atomic number Z. %(i.e. the number of electrons in the atoms) up to a saturation threshold [].  
For deliberate heating strategies, the potentially longer absorption lengths enable more homogenous heating compared, e.g. to optical lasers or ion beams\cite{Nagler:2009,Levy:2015,Saunders:2018},
%USED NaglerNatPhys09
 and scaling up of targets to enable larger irradiated volumes\cite{Levy:2015,Saunders:2018}.  %X-rays enable extremely tight beam foci, and hence high irradiance.  
X-ray heating performed with %intense x-rays generated by 
large optical laser\cite{Ping:2015,Saunders:2018}, 
%USED Ping15; 
%preheating papers
%Z-pinch
pulsed-power\cite{Bailey:2003}, 
%USED http://www.sciencedirect.com/science/article/pii/S0022407303000505
and %$%soft x-ray 
free electron laser (FEL) \cite{Zastrau:2008,Glenzer:2016,Nagler:2009,Zastrau:2014,Ping:2015,Fletcher:2014,Levy:2015}
%see below and FIX IN TABLE: 
%USED Nagler,
%USED Zastrau, 
%USED Vinko 
%etc.: 
%moveing main list here: Vinko2012, Nagler09, Zastrau14, Ping15, 
%USED Fletcher:2014 : http://aip.scitation.org/doi/pdf/10.1063/1.4891186, 
%USED Glenzer:2016, 
%[\textit{check that these are softer; fix values in table}]
facilities have been demonstrated %largely
extensively at lower photon energies (hundreds of eV to several keV) which can
%similar to other lower photonic energy experiments are limited in 
still limit the potential thickness and materials of target components %, including tampers,
 and hence %in
  experimental timescales.%; s often comprise divergent sources [] which limit experimental geometries and brightness [
%See also http://iopscience.iop.org/article/10.1088/0953-4075/49/9/092001/pdf summarises laser based sources
%].

Free electron lasers and other high-brightness x-ray sources operating in the hard x-ray regime above $\sim$10 keV  (Table \ref{tab:facilities}) allow for substantial scaling up of target dimensions and experimental timescales. At x-ray energies exceeding $\sim$10 keV, absorption lengths in even heavy-element solids exceed several $\mu$m enabling large volume homogenous irradiation\cite{Levy:2015,Saunders:2018}. % (e.g. over 0.5 $\mu$m in Ag at 8.9 keV).  
Moreover, x-ray absorption lengths are at the $\sim$mm level above 10 keV in common light element solids, allowing delivery of x-ray energy through thick low-Z tampers to high-Z samples.  The possibility of massive tampers %, orders of magnitude thicker than samples, 
which remain cold and stable during the experiment, and which completely control the sample's expansion, 
may thus be realized with such hard x-ray sources.  %
For hard x-ray FELs, the high total pulse energy ($\sim$1 mJ, or 10$^{12}$ photons), fast timescale (10-100 fs), and high brightness ($\sim$10$^{18}$ W/cm$^2$) is comparable to typical optical laser systems;
%while x-ray energies exceeding 10 keV can more readily penetrate thicker targets.  
similar total energies in somewhat longer pulses ($\sim$100 ps) are possible at fourth-generation synchrotron radiation sources %, achieving comparable pulse energies in the hard x-ray regime but are also forthcoming 
(Table \ref{tab:facilities}).  In addition to presenting challenges in adapting conventional x-ray probing studies to modern brilliant light sources, these capabilities presage a new generation of irradiative extreme temperature experiments. Radiatively heated samples in such experiments can, depending on target design, survive longer than those in lower energy experiments, enabling the achievement and exploration of more nearly thermal and thermodynamic equilibrium conditions, and study of processes normally out of range in ultrafast experiments such as diffusive heat conduction, bulk chemical reaction, equilibrium phase transformation, and phase separation and mixing.
% larger scale of  targets.  and confinement and longer experiments in the equilibrium limit.  
% This enables %new kinds of 
%heating %strategies, operating 
%strategies approaching the LTE limit.  

%%%%%%%%%%%%%%%%%%%
\begin{table*}[!htb]
\begin{center}
\centering
\footnotesize\setlength{\tabcolsep}{5pt}
\begin{tabular}{|c|c|c|c|c|c|}
\hline 
%\multicolumn{6}{|c|}{X-ray Sources} \\\hline
%Medium & \multicolumn{3}{|c|}{Materials} & Photon  & Energy \\
 %      Thickness[$\mu$m] & Foil & Medium & Tamper &  Energy[$ke$V] &  Pulse[$\mu$J] \\ \hline \hline   
 %      Thickness[$\mu$m] & Foil & Medium & Tamper &  Energy[$ke$V] &  Pulse[mJ] \\ \hline \hline 
%\textit{5} &  \textit{Fe} &  \textit{Al$_{2}$O$_{3}$} &  \textit{Diamond} &  \textit{25} &  \textit{0.35} \\ \hline \hline
\multicolumn{6}{|c|}{4$^{th}$ Generation X-ray Sources} \\ \hline
%Sec1 &	Sec2	&Sec3             &	Sec4	&Sec5	&Sec6\\\hline

%Facility & \multicolumn{3}{|c|}{Pulse} & Photon  & Energy \\
%       Thickness[$\mu$m] & Foil & Medium & Tamper &  Energy[$ke$V] &  Pulse[$\mu$J] \\ \hline 
%       Thickness[$\mu$m] & Foil & Medium & Tamper &  Energy[$ke$V] &  Pulse[mJ] \\ \hline 
%Facility & \multicolumn{3}{|c|}{Pulse} & Photon  & Energy \\
 & \multicolumn{2}{|c|}{Pulse} &   X-ray &  Minimum & Pulse \\
 Facility & Duration & Energy & Energy & Spot Size & Delay \\ 
   & [ps] & [mJ] & [keV] & [$\mu$m]  & \\ \hline 
\multicolumn{6}{|c|}{Hard X-ray Free Electron Lasers} \\ \hline

%LCLS\cite{MECweb} & 0.08 & 1-3 & 3-12 & 3  &  8.3ms  \\ %\hline %1.1\times10^[17} intensity, but from where?
LCLS-II-HE\cite{MECweb, LCLSII, LCLSIIHE} & 0.01-0.06 & 1-3 & 25 (12.8) & 3 & 8.3ms (1$\mu$s) \\ \hline
%pulse energy .15*10^12*25000*1.60218e-19, after https://lcls.slac.stanford.edu/sites/lcls.slac.stanford.edu/files/LCLS-parameters-07-16-19.pdf
%ZUZANA SAYS 1 us rep rate at LCLS II is not possible at 25 keV?
European XFEL & 0.05 - 0.1 & .35 - 4 & 5 - 20 & $<1$ & 220ns \\ \hline
SACLA\cite{Yabashi:2015} & 0.01 &  0.5 & 4-15 & 1 & 17ms  \\ \hline
\multicolumn{6}{|c|}{Synchrotron Upgrades} \\ \hline
ESRF-EBS\cite{Garbarino:2017} & 100 & .04 & 10-70 & 0.15 & 176ns  \\ \hline
%PETRA IV & & & & &  \\ \hline
%APS-U & & & & &  \\ \hline
%\textbf{[APS should upgrade to 100x intensity in 2022.  Get this reference. Answer: could not find relevant on target parameters & very early paper suggests only 10x in ph. per pulse ]}
%\multicolumn{8}{|c|}{Optical Lasers} \\ \hline
%Omega & & & & N/A & & & Divergent \\ \hline
%
%\multicolumn{7}{|c|}{Soft X-ray / UV Free Electron Lasers} \\ \hline
%FLASH & 0.01-0.2 & 1 & 0.015-0.3 & 5 & $10^{17}$ & 1 $\mu$s  \\ \hline
%LCLS-SXR & & & & & &  \\ \hline
%
%\multicolumn{8}{|c|}{Z-pinch} \\\hline
%? & & & & N/A & & & Divergent \\ \hline
 %\multicolumn{8}{l}{A:  Note: }\\
 \end{tabular}
\caption{Comparison of typical operating parameters of pulsed, focused x-ray facilities, with representative first-harmonic capabilities of current-generation XFELs and a representative 4th generation synchrotron upgrade.}
\label{tab:facilities}
\end{center}
\centering
\end{table*}
%%%%%%%%%%%%%%%%%%% 

%Brilliant x-ray irradiation from current and emerging central facilities can thus lead to significant energy deposition in targets in the fs-ns timescale %[\textit{see above, select harder x-ray examples}]
%, with one result being thermal energy deposited differentially in the target %USED Ping15;?
%Due to differential energy deposition in targets depending on local x-ray absorption behaviour\cite{Ping:2015}.
% g in x-ray absorption cross sections and the high transparency of lower-Z matter to hard x-rays, substantial energy deposition differences in light element (low-Z) and heavy element (high-Z) materials 

%suggest particular target designs using tamped configurations that can be used to control the effects of irradiation.  Specifically, 
%A tamped configuration would comprise a target with a high x-ray absorption sample and a tamper with a relatively lower one, practically implying a higher-$Z$ sample and a lower-$Z$ tamper.   

Many interesting and poorly understood phenomena at warm dense matter conditions are found at elevated densities, i.e.  exceeding that of the solid state, including metallization of molecular insulators\cite{Celliers:2018} and phase separation in warm dense mixtures \cite{Kraus:2017,McWilliams:2015,Schottler:2018}.  To access these conditions via irradiative heating requires that samples be initially pre-compressed to the needed density.  The effects of increasing density on fundamental interactions in irradiatively heated matter including bonding\cite{Medvedev:2013, Johnson:2005} and electron-ion thermalization\cite{Zastrau:2014,Nagler:2009,White:2014} also require investigation.
%, such as fluid-fluid phase transitions
%, phase separation of noble gases
%, and metallization in molecular materials
%\cite{Schottler:2018} []
%add Celliers%CelliersScience
%Lorenzen PRL 2009
%USED (Zastrau PRE 2014; http://journals.aps.org/pre/pdf/10.1103/PhysRevE.90.013104, 
%USED NaglerNatPhy09, 
%[
%all EIE
%].
%The laboratory exploration of matter at conditions of extreme temperature, density, and pressure are needed to understand natural systems from stellar and planetary interiors [PNAS15, ?] and technological systems including inertial confinement fusion.  
The ability to employ confining tamper layers of substantial thickness in hard x-ray experiments (if of sufficiently low-Z composition) raises the possibility of using these layers as anvils to apply initial pressure to matter prior to x-ray probing or excitation. Such a design is commonly used in static high pressure devices, notably the diamond anvil cell (DAC), which employs thick (several mm) diamonds to isothermally compress thin samples to %produce very 
high pressure and density %states in samples prior to irradiation
\cite{Mao:2007}.  Long-used at synchrotron facilities, and compatible with hard x-ray illumination as either a %non-intrusive 
probe %[Some Synch rad paper, Mao\&Mao07] 
or %as a 
pump, %to alter [samples deliberately/sample properties] 
%[H2O paper Mao Science?]
the DAC offers the possibility to study %high energy-density 
the properties and dynamics of high density, pressure and temperature material states on ultrafast timescales when coupled to brilliant x-ray sources.  
Many x-ray measurements developed for static high-pressure devices at traditional synchrotrons stand to be adapted for use at modern higher-brightness sources, such as characterization of dynamic pressure and temperature modulation\cite{Jenei:2019,Goncharov:2010} with serial x-ray probing (Table \ref{tab:facilities}).
%\cite{\McWilliams:2015, McWilliams:2016,Garbarino:2017,Goncharov:2012,Dubrovinsky:2012,Jeanloz:2007,Armstrong:2010,Mao:2007}
Static compression can also maintain sample confinement and high density during heating to the electron-volt ($>$10,000 K) temperatures of warm dense matter\cite{McWilliams:2015}, allowing near-isochoric experiments orders of magnitude beyond hydrodynamic timescales.

The purpose of this study is several-fold, and motivated by the increasing brightness of hard x-ray sources
%s from free electron lasers, upgraded synchrotrons, and laser driven 'backlighter' x-rays from large laser facilities. These can 
providing fast pulsed (nanosecond to femtosecond) hard x-rays (to tens of keV) at high power (10$^{11}$-10$^{12}$ photons per pulse).  The main objective is to explore the thermal and mechanical evolution of pulse-irradiated targets involving particularly thick tampers, a configuration suggested by the ability of hard x-rays to pass unimpeded through low-Z tampers to a high-Z target layer confined within, to which energy is delivered.  One application of interest is extending isochoric radiative heating studies %to the equilibrium limit
%   limit in which electrons and ions are typically out of equilibrium and experiments are cut off by hydrodynamic expansion of thin targets.  B
by delaying or inhibiting altogether hydrodynamic expansion, so that matter can be observed at thermal, and plausibly thermodynamic, equilibrium while at extreme temperature and near-solid density.  
%This would be invaluable for making observations of warm, dense matter properties of direct relevance to materials phase diagrams, planetary science, and other systems involving near-equilibrium plasmas, solids, and liquids, including inertial confinement fusion.  
%In contrast to traditional isochoric heating, on these larger timescales and lengthscales LTE can be assured and target conditions develop as a result of conventional hydrodynamic processes and diffusive heat transport as explored in the present modelling.
%Thus, there is a need to examine the types of energy deposition, confinement, excitation, and relaxation that may occur in materials and targets under this kind of irradiation. 
%One persistent question regards the degree of heating effected by x-ray irradiation\cite{Mo:2018}, and this study highlights phenomena that, if measured, could constrain these effects.  %These processes are generally well understood for ultrafast experiments, as discussed above for optical laser methods[], but such experiments typically end within a nanosecond.  Here we examine processes relevant to longer duration experiments, as are [increasingly] made possible by intense hard x-ray sources which enable experiments on bulk targets of larger physical size.
A related objective is to characterize the performance of diamond anvil high-pressure cells (DACs), long used to great effect in synchrotron x-ray science, at higher intensity pulsed x-ray sources 
%%USED McWPNAS,GonchXRAY
where heating during the x-ray exposure could be an unavoidable %, but controllable,
byproduct of %using these high brightness 
x-ray probing or used deliberately to heat pre-compressed matter to extreme temperature, as %from another perspective,
an alternative to %standard 
optical laser heating\cite{Dewaele:1998,Mao:2007, Goncharov:2010,McWilliams:2015}. The response of the anvil-cell type of tamped target to high brightness irradiation, and the designs it inspires for general tamped laser-matter interaction experiments, are discussed in Sec. \ref{sec:DAC}.
We also aim to characterize in general the heat dissipation in solid layered targets which may be of practical use as beamline optics\cite{Sinn:2007} and detectors\cite{Roth:2018} at x-ray facilities. The survival of these components often depends on their heat and stress dissipation capabilities and often utilize high strength, high thermal conductivity materials such as diamond\cite{Sinn:2007,Roth:2018}.%, %[
%aboverefs
%], 
%a material examined here.

\section{\label{sec:methodology} METHODOLOGY}

Targets simulated here consist of a sample layer or layers ($\mu$m thickness) between thick (mm thickness) tampers.  The advantages of this configuration are: (1) exceptionally long confinement of samples at extreme conditions, so that the approach to, and properties of, thermodynamic equilibrium states of high density and temperature can be studied; (2) efficient control of sample temperature by using high thermal conductivity tampers, enhancing sample stability and promoting sample survival after irradiation; and (3) the ability to pre-compress samples with strong tampers, and resist thermomechanical stresses developing during the irradiation.  %A tamped sample configuration of particular interest comprises two thick diamond tampers with a thin sample contained within, the configuration of a diamond anvil cell, which is discussed in detail in Sec. \ref{sec:DAC}.

The thermomechanical response of these micron-to-millimeter scale x-ray heated layered targets evolves on a range of timescales.  We consider a high-brightness monochromatic hard x-ray source, with a FEL-like beam diameter and pulse duration, delivering heat energy by x-ray absorption in $\sim$ 100 fs.  Pressure waves generated by thermal expansion propagate on ps-ns timescales, adiabatically mediating pressure and temperature evolution in the differentially heated target; the timescale\cite{GomezPerez:2017} is set by the scale length of the heated volume $\ell$ divided by the sound speed $c$, i.e. $\ell/c$.  Adiabatic conditions break down on ns-$\mu$s timescales, with heat conduction cooling heated areas toward the initial temperature, at which the surrounding target remains; the timescale\cite{GomezPerez:2017} of this process is roughly the square of the heated volume size divided by the thermal diffusivity coefficient $\kappa$, or $\ell^2/\kappa$.  On these lengthscales (micron to millimeter) and timescales (ps and longer) LTE can be assumed, and target conditions develop primarily as a result of conventional hydrodynamic processes and diffusive heat transport in locally equilibrated matter;
%as explored in the present modelling
 near-isochoric conditions are assumed to be maintained throughout by stable tampers.   %This contrasts with isochoric heating performed on thin and/or unconfined targets, where equilibration dynamics overlap with hydrodynamic and heat-transfer processes and experimental timescales are often limited in the ps range.  
 %Ultrafast processes occurring during and immediately subsequent to energy delivery 
%The approach to LTE on timescales in the range of a few ps after the irradiation is not included; iss
%In the present limit of massive samples and time and length scale, equilibration processes occur rapidly and locally.     We are primarily interested in the subsequent effects in the target once LTE has been established.  This includes the

To study heat conduction, we use a two-dimensional finite element (FE) model including conduction along and lateral to the x-ray beam path, both important on the associated ($\mu$s) timescales for tightly focussed radiation (Sec. \ref{Sec:ModelsFE}).  To study the hydrodynamic processes, which can take the form of shock discontinuities, we separately employ one-dimensional radiation hydrodynamics models to study the mechanical and associated thermal evolution of the system for the first few ns (Sec. \ref{Sec:ModelsHydro}); this approach is chosen because finite element models are not well suited to stress waves of larger magnitude, and because, if beam diameter is kept greater than the thickness of the relevant layers, the initial evolution of sample conditions is accurately treated as a one-dimensional process in the direction of the beam.

\subsection{Finite Element Models}\label{Sec:ModelsFE}

\subsubsection{General approach}

In order to describe the pulsed x-ray heating and cooling of a tamped sample configuration, we used a simulation software (\textsc{comsol} \footnotesize{Multiphysics}\normalsize) based on finite element analysis to implement a two-dimensional, time-dependent heat 
%and radiative 
transfer model\cite{Konopkova:2016, Goncharov:2012, GomezPerez:2017, McWilliams:2015}, with semi-transparent materials exhibiting a bulk absorption of the x-ray radiation.
We simulate the case of a single intense x-ray pulse of $\sim$100 fs duration, and later (Sec. \ref{Sec:PulseTrain}) a train of such pulses, striking a sample initially at room temperature (300 K). %  These simulations are later extended to describe the effects of pulse train heating. % and can be adjusted and pulses of different lengthscales and diameters.

%%%%%%%%%%%%%%%%%%
\begin{figure}
    \centering
    \includegraphics[width=8.5 cm]{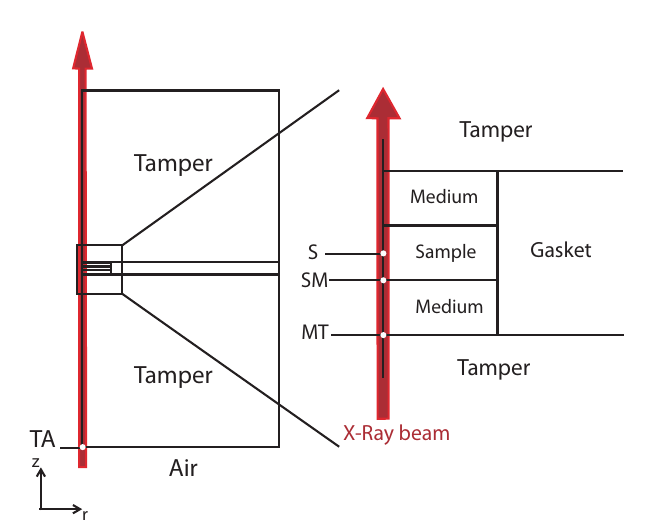}
    \caption{Schematic illustration of the general model geometry, depicting the axis-symmetric slice from the axis to the edge of the cylinder.  For finite element models, a 2D cylindrical geometry 160 $\mu$m in radius and 4005-4025  $\mu$m in length is employed.  For hydrodynamic models a simple 1D representation of the boxed region is used.  X-rays are incident from below.  Standard dimensions are specified in Table \ref{tab:parameters}. %The dashed region represents the volumetric zone on the incident beam pass where the Beer-Lambert law is applied and the only region that the beam is heating up. 
   Measurements are taken at S (sample center), SM (sample-medium interface), MT (medium-tamper interface), and TA (tamper-air interface), with interfaces referring to the leading (upstream) interface unless otherwise indicated.  }
    \label{fig:schemaDAC}
\end{figure}
 %%%%%%%%%%%%%%%%%%
%(\cref{fig:DAC}).  
%%
%\begin{figure}[htbp]
%	\centering
%	\includegraphics[width=8.5 cm]{../figures/exportfigures/Fig01} 
%	\caption{This figure shows the schematic configuration of the modelled domain. We assumed a cylindrical symmetry around the axis shown in purple. The coupler disk (orange) is made of an optically opaque material and is heated with lasers on surfaces $s_1$ and $s_2$, shown in green. Inside the sample cylinder (white) the pressure medium is optically transparent and it is heated only by the heat transferred from the interior disk. Outer boundaries of the sample chamber are kept  at a constant temperature of \SI{300}{\kelvin}.  
%	}
%	\label{fig:DAC}
%\end{figure}
%%

%%%%%%%%%%%%%%%%%%%%%%%%%%%%%%%%%%%%%%%%%%%%%%%%%%

Assuming a multilayer target of layers perpendicular to the incident x-ray beam (Fig. ~\ref{fig:schemaDAC}), we exploit the symmetry around the beam, and consider a two-dimensional model by a rotational symmetry about an axis through the center of the beam path, with $z$ referring to the axial position and $r$ the radial position.  %By reducing the model to a 2D axisymmetric geometry, simulation times are minimized. 
The pulsed x-ray beam propagates in the $+z$ direction, centered at $r=0$. 
Including time $t$, this model is three dimensional.   We vary the geometries of the layers used in the system as needed to simulate different configurations. Thick, low-Z tampers (or anvils) of 2 mm thickness are placed on either side of a primary sample `foil' layer of 5 $\mu$m thickness.  Additional interfacial layers (or medium), of several $\mu$m thickness, are used between the tamper and foil in most simulations.  The medium can play several roles in experiments, acting as: (1) a protective layer, preventing direct heating of the tamper and absorbing thermal stress when resisting hydrodynamic expansion; (2) as an insulating layer to extend the experimental duration by limiting cooling of the sample; and (3) as a hydrostatic pressure medium, in cases where the target is configured as a high pressure cell.  The sample (and where used, medium) are contained laterally by a thick layer bridging the two tampers (or gasket, a component designed to reflect the configuration of anvil cells, which has little effect on the simulations). Typical dimensions are shown in Table \ref{tab:parameters}.   This geometry is also symmetric about a parallel plane through the middle of the sample layer; conditions achieved, however, are asymmetric about this plane.
Constant volume conditions are assumed, which is appropriate if targets remain in the condensed state (i.e. below vaporization points) or where they are configured to resist thermal stresses and hydrodynamic expansion, e.g. using thick tamper layers or an anvil cell design having a fixed sample cavity volume\cite{Dewaele:1998}. The effects of thermal expansion and stress waves are treated separately as these occur on significantly different timescales and require a self-consistent hydrodynamic approach due to the rapid nature of heating and consequent shock production (Sec. \ref{Sec:ModelsHydro}).%; related stresses relax in $\sim$$10^{-9}$ s, before the heat conduction treated in these models begins.

%%%%%%%%%%%%%%%%%%%%%%%%%%%%%%%%%%%%%%%%%%%%%%%%%%%%%%%%
 \begin{table}[htbp]
\begin{center}
    \begin{tabular}{ |c|c|} 
 \hline
 \multicolumn{2}{|c|}{Fixed Dimensions, } \\ 
 \multicolumn{2}{|c|}{ Finite Element Models} \\ \hline
    Parameter & Value[$\mu$m] \\ \hline
 Model Domain Radius& 160 \\ 
  Foil Radius & 40\\ 
  Medium Radius   &40\\ 
 Tamper Thickness ($d_T$) & 2000  \\ 
 Foil Thickness ($d_S$) & 5 \\ \hline
   \end{tabular}
 \caption{Geometric constant parameters for finite element modeling.} 
 \label{tab:parameters}
\end{center}
\end{table}
%%%%%%%%%%%%%%%%%%%%%%%%%%%%%%%%%%%%%%%%%%%%%%%%%%%%%%%%

%Laser heating of solids can be modelled as a simply as statement of the law of conservation of energy in differential form.  The relationship between the heat flux or the thermal energy crossing unit area per unit time $J(r,t)$ and the temperature $T(r,t)$ is given by Fourier's Law[]
%%\textbf{REFERENCES}] 
%\begin{equation}
%J(r,t)= -k \nabla T(r,t),
%\end{equation}
%where $k$ is the materials thermal conductivity. 

In order to describe the dynamical temperature evolution of targets, % under various heating conditions,
we used the finite-element solution of the time-dependent energy transfer equation.
The volumetric heat source $Q(r,z;t)$ (the net energy generated per unit volume and time) representing the radiative energy absorbed within the target is given as 
%%%%%%%%%%%%%%%%%%%%%%%%%%%%%%%
\begin{equation} 
\label{eq:heattransfer}
%Q(r,z;t)=\rho C_P \frac{\partial{T(r,z;t)}}{\partial{t}} + \nabla \cdot (-k \nabla T),
Q(r,z;t)=\rho C_P \frac{\partial{T}}{\partial{t}} + \nabla \cdot (-k \nabla T),
\end{equation}
%%%%%%%%%%%%%%%%%%%%%%%%%%%%%%%%
%$k$ the thermal conductivity, and $Q$[W/m$^{3}$] is the net energy generated per unit volume and time or simply volumetric heat source. 
%After energy is deposited, and assuming physical properties such as $k$ are constant this equation takes the form
where $T$ is the temperature, $t$ is the time, $k$ is the thermal conductivity, $\rho$ is the density, and $C_{P}$ is the heat capacity at constant pressure.  For constant physical properties, and considering the period after heating, Eq. \ref{eq:heattransfer} reduces to 
\begin{equation}
\frac{\partial T}{\partial t}=\kappa\nabla^2T,
\label{eq:mainpartial}
\end{equation}
where $\kappa$ is the thermal diffusivity,
\begin{equation}
\kappa=\frac{k}{\rho C_P}.
\end{equation}
%where ,. [\textit{disconnect here...Natalia?}]
%At the surfaces of strongly emissive materials, such as metallic samples, the temperature rises modestly and radiative heat loss through transparent adjacent layers can occur.
%, however due to the large temperature gradients on the surface $(z=0)$ material assuming equal to the thermal radiation loss in the two faces of the cylindrical geometry and the reflective properties for each material interfaces. The radiation loss reads as Equation[~\ref{eq:radiationloss}]. 
% %
%%%%%%%%%%%%%%%%%%%%%
%\begin{equation} 
%\label{eq:radiationloss}
%k\Eval{\frac{dT(r,t)}{dz}}{r=0}{}= \sigma \epsilon_{s}
%\left[T^{4}(z=0,t)-T_{\infty}^{4} \right],
%\end{equation}
%%%%%%%%%%%%%%%%%%%%%%%%%
%Where T(r,t) is the target temperature, $k$ is thermal conductivity, $\sigma$ is the Stefan-Boltzmann constant, and $\epsilon_{s}$ is the surface emissivity.
%As the large temperature gradients due to incident laser intensity and the 
%However, 
Radiative (photon) heat transfer is generally negligible compared to diffusive (phonon and electron) heat conduction at the presently examined temperatures and timescales\cite{GomezPerez:2017}, and is not included. 
%in the calculations.
%[Dewaele et al., 1998; Manga and Jeanloz, 1997; Montoya and Goncharov, 2012, Geballe 2012, McWilliams PEPI 15].
%(Equation[~\ref{eq:radiationloss0}]. ([\textbf)
%
%
%%%%%%%%%%%%%%%%%%
%\begin{equation} 
%\label{eq:radiationloss0}
%k \frac{dT(r,t)}{dz}|_{r=0} = 0,
%\end{equation}
%%%%%%%%%%%%

The source term $Q(r,z;t)$ (typical units of W/m$^{3}$) is given by volumetric heat generation when the incident x-ray beam passes through, and is absorbed within, the semi-transparent materials.  Due to this absorption the beam intensity decays exponentially with depth (Lambert-Beer law).  At the considered x-ray energies, the contribution of diffuse scattering to total attenuation is small %[website]
 and is neglected in our calculations.  Coherent scattering (i.e. Bragg diffraction) could become important particularly where thick single crystals are used as tampers, affecting attenuation and radiation trajectory, though, as it can be avoided in practice\cite{Loveday:1990}, it is also ignored. %; Bragg contributions to attenuation and associated redirection of beam energy  %Such Bragg diffracted beams could be both intense and localised, causing heating as well as probing at unexpected regions of targets adjacent to the column/area of direct irradiation.  
 %At interfaces, this beam can also be partially reflected backwards.  
 %, with a smaller fraction due to scattering <and other processes>.  %For example compton scattering? More serious for lower Z sample.
%X-ray attenuation is presumed to be due entirely to absorption.  This is appropriate at the presently examined energies (?) as scattering is generally much weaker than absorption [cite tables used to determine cross sections].  
% in further discussion of our technique[, though representative values of $R$ were included in our simulations]. 
%The attenuation of x-ray intensity follows the Lambert-Beer law, and this also gives the local source term for bulk photo-absorption and heat generation in $Q(r,t).
%A normal 
%X-Ray beam reach the surface and reflectance was considered at each interface at the specific photon energy simulation see Table[~\ref{tab:physicalproperties}].
%%%%%%%%%%%%%%%% Dr McWilliams
%\medskip 
The energy deposition in a given homogenous layer in a target can thus be written as %given by Eq. [~\ref{eq:attenuation}]
%%%%%%%%%%%%%%%%%%%%%%%%%%%%%%%%
\begin{equation} 
\label{eq:attenuation}
Q(r,z;t)= I_s(r;t) \alpha(1-R_s)\exp[{-\alpha (z-z_s})]
\end{equation}
%%%%%%%%%%%%%%%%%%%%%%%%%%%%%%
where $\alpha$ is the absorption coefficient, constant in the layer, $z_s$ is the $z$ position of the layer surface the radiation is incident on, $R_s$ is the reflectivity of the leading surface or interface, and $I_s(r;t)$ is the incident intensity on the surface (typical units of W/m$^{2}$).  For x-ray radiation, reflectivities of interfaces are exceedingly small, of order $R_s\sim 10^{-9}-10^{-13}$, and may be neglected.
Thus the attenuation of x-rays as well as the energy deposition is accurately estimated by considering absorption only.

The absorption in the target is given by computing the sequential absorption in several such layers. At the downstream surface of a layer, % the incident beam direction,
 boundary conditions establish that any light reaching that boundary will leave the domain and pass to the next layer and this is repeated until the beam reaches the downstream target surface and leaves the geometry. For example, in the center of the sample (and target), we have %Eq. [~\ref{eq:attenuation2}]
%%%%%%%%%%%%%%%%%%%%%%%%%%%%%%
\begin{eqnarray} 
\label{eq:attenuation2}
&Q(r=0,z=z_c;t)=\nonumber\\
&I(r;t)\alpha_{S}\exp({-\alpha_{S} \frac{d_{S}}{2}})\exp({-\alpha_{M} d_{M}}) \exp({-\alpha_{T} d_{T}}),
\end{eqnarray}
%%%%%%%%%%%%%%%%%%%%%%
where $S$, $M$, and $T$ refer to the sample, medium, and tamper values, respectively, $I(r;t)$ is the incident intensity on the target assembly, $d$ refers to the thickness of particular layers, and $z_c$ refers to the center of the sample layer (and target assembly), hence only half of the sample's thickness is included.

%%%%%%%%%%%%%%%%%%%%%%%%%%%%%
\begin{table}
\begin{center}
\begin{tabular}{ |c|c|} 
 \hline
 \multicolumn{2}{|c|}{Pulse Parameters,} \\
 \multicolumn{2}{|c|}{Finite Element Models} \\
 \hline
  Parameter & Value[units] \\ \hline 
      Arrival time ($\mu$)  & 400[fs] \\ 
  Pulse length ($\sigma_t$)  & 100[fs]  \\ 
   Pulse size ($\sigma_r$)    & 5[$\mu$m]\\ \hline 
      \end{tabular}
 \caption{Parameters for the x-ray pulse in finite element models.} 
 \label{tab:ParametersGauss}
\end{center}
\end{table}
%%%%%%%%%%%%%%%%%%%%%%%%%%%%%%%%%%%%%

The model considers heating induced during a $\sim$100 fs duration x-ray pulse, and the conductive heat transfer following the rapidly imposed temperature distribution in the target. The heating pulse intensity is assumed to follow a Gaussian distribution in time and space, with incident intensity $I(r;t)$  (Eq. ~\ref{eq:Intensity}) reaching a maximum, $I_{max}$, at $t=\mu$ and $r=0$ as
%%%%%%%%%%%%%%%%%%%%%%%%%%%%%%%%%%%%%%%%%%%%%%%%%%%%%%%%%%%%%%%%%
\begin{equation} 
\label{eq:Intensity}
I(r;t)=I_{max}\exp\left[{-\frac{r^2}{2\sigma_r^2}}\right]\exp\left[{-\frac{(t-\mu)^2}{2\sigma_t^2}}\right],
\end{equation}
%%%%%%%%%%%%%%%%%%%%%%%%%%%%%%%%%%%%%%%%%%%%%%%%%%%%%%%
where $\sigma_r$ is a Gaussian radius parameter, such that the FWHM (full width at half maximum) diameter of the pulse is 
\begin{equation} \label{Eq:spotsize}
\text{spot size}=2\sqrt{2 \ln 2} \sigma_r, 
\end{equation}	
and $\sigma_t$ defines the temporal width of the pulse (FWHM) as 
\begin{equation}
\text{pulse duration}=2\sqrt{2 \ln 2} \sigma_t.
\end{equation}
 %the spot [diameter/radius]  and pulse width respectively, defining when the power raising at $\mu$ (Pulse Time) of the simulation. 
 For the parameters of this simulation (Table \ref{tab:ParametersGauss}) the spot size is then $\sim$12 $\mu$m, and the pulselength $\sim$240 fs. %(Table~\ref{tab:ParametersGauss}).
The incident peak intensity $I_{max}$ can be related to the net energy of the single pulse $E_{pulse}$ (in J), the peak incident power $P_{max}$ (in W, and occurring at $t=\mu$), and the peak energy density per area $\Lambda_{max}$ (in J/m$^2$, and occurring at $r=0$)  %[\textit{Stewart: check this \& where used to compare to Medvedev damage threshold}] 
as
\begin{align} 
 \label{eq:PowerEnergy1}
I_{max} &=\frac{E_{pulse}}{(2\pi)^\frac{3}{2} \sigma_t\sigma_r^2}\\
% &= \frac{P_{max}}{\pi \sigma_r^2} \\
\label{eq:PowerEnergy2}
 &= \frac{P_{max}}{2 \pi \sigma_r^2} \\  
% & \text{POSSIBLE ERROR ABOVE, MABYE BELOW}\\ factor of two fixed
% &= \frac{\Lambda_{max}}{2 (2\pi)^\frac{1}{2} \sigma_t }
% \label{eq:PowerEnergy2}
\label{eq:PowerEnergy3}
&= \frac{\Lambda_{max}}{ (2\pi)^\frac{1}{2} \sigma_t }  
% \label{eq:PowerEnergy3}
\end{align}
The number of photons per pulse $N$ is
\begin{equation}
N=\frac{E_{pulse}}{E_{photon}}
%&=\frac{E_{pulse} \lambda}{hc}
\label{Eq:Nphotons}
\end{equation}
and is equivalent to $\sim$$10^{12}$ for the peak energy per pulse (3.5 mJ) and x-ray energy (25 keV) simulated here.  In our models we specify $E_{pulse}$ (Eq. \ref{eq:PowerEnergy1}), which when integrated over the pulse duration (Eqs \ref{eq:attenuation2} and \ref{eq:Intensity}) leads to $Q(r,z;t>>\mu)$ independent of the pulse duration, such that $T(r,z)$ immediately after the pulse (and before significant heat transport occurs) depends only on total pulse energy and its spatial distribution, i.e. temperature achieved is independent of pulselength so long as the pulselength is shorter than heat conduction timescales. This implies any pulse duration less than the heat conduction timescales (roughly in the ns range or less) will achieve similar peak temperature and show identical cooling behavior.  %A more detailed treatment of heating, discussed below, might include additional phenomena dependent on the pulse intensity and duration.

%The geometry and computational assumptions are shown in Table[~\ref{tab:parameters}].  

%The incident intensity shown in Equation[\ref{eq:attenuation}] according on cylindrical geometry $I(r,t)$ times the absoption became the $Q(r,t)$ this term could expressed in energy of the pulse values and the energy of the heat source and its reads  in Equation(~\ref{eq:PowerEnergy})

 %%%%%%%%%%%%%%%%%%%%%%%%%%%%%%%%%%%%%%%%%%%%%
 
The initial temperature of the entire system is assumed to be ambient (300 K).
%The simulations are initialized with all the cavity at ambient temperature (300 K).  
As a boundary condition, the external surface of the simulation cell %target %tamper
%[\textit{John: what about gasket?}] 
shown in Fig. \ref{fig:schemaDAC} was given by natural heat exchange with a surrounding atmosphere (air), with the external temperature fixed at 300 K, and heat loss from the surface determined %[\textit{John: what reference?}]
%\textbf{references} 
as
%], represented as Equation[~\ref{eq:convection}]
%%%%%%%%%%%%%%%%%%%%%%%%%%%%%%
\begin{equation} 
\label{eq:convection}
q_0=h(300 \text{ K}-T)
\end{equation}
%%%%%%%%%%%%%%%%%%%%%%%%%%%%
where $q_{0}$ is the convective heat flux and $h$ is the convective heat transfer coefficient ($h=5$ W/m$^{2}$/K, for natural convection in air). %considering a 300 K temperature outside the geometry.  
This has no significant effect for the cooling timescale of these experiments; similar results could be expected in vacuum.
%[On the timescale of these experiments, this effect has a very small effect and so the total sample a]
%The exterior of the cell is presumed to be in contact with air and loses heat via convective heat loss.

A free triangular mesh is employed, which is kept very fine at interfaces due to the need to stabilize the model during the initial phase of large temperature gradients at interfacial regions, at heating times $10^{-12}$ to $10^{-9}$ s; the heat transfer starts at approximately on $10^{-9}$ s time scales, and temperature is stable before this if the simulation is configured properly.  A coarser mesh is used away from the interfaces.
The accurate modeling of interfaces on shorter timescales is validated analytically (Sec. \ref{sec:GeneralObservations}).
%and required especially fine zoning of the simulation near the interface; if this was not performed, significant anomalies in temperatures with unphysical rapid variations were produced.

As the simulations seek to establish general trends for the effects of target composition, geometry, and beam parameters, a number of physical assumptions are made in our calculations.
We assume a direct relationship between the amount of x-ray energy deposited in the target at a given location and the amount of heating at this location.  Further, the models assume that thermal equilibrium (i.e. between electrons, which initially absorb energy, and ions, which heat more gradually on the ps timescale of electron-ion equilibration)
%Covered Earlier
%\cite{White:2014,Ping:2006,Ping:2015} 
%[WhitePRL14, Gold, Ping papers, Celliers, select from above]) 
occurs instantly. Thus our simulations should be accurate at timescales of $\sim$10$^{-12}$ s and longer, i.e. sufficiently greater than electron-ion equilibration times, while only approximating the initial (fs) heating process.  Implicitly, we also assume localization of hot electrons during the equilibration period, i.e. that any hot electrons produced ultimately equilibrate with nearby ions.  This is a reasonable approximation since the typical mean free path of ballistic hot (eV) electrons in condensed matter tends to be of order $10^{-2}$ $\mu$m \cite{Ping:2015,Nagler:2009,Medvedev:2013}, which is much less than the sample dimensions and heating beam diameter (1-10$^3$ $\mu$m), consistent with a diffusive heat transfer model being sufficiently accurate on these time and lengthscales.  While not included here, hydrodynamic (Sec. \ref{Sec:ModelsHydro}) and radiative processes, longer-distance hot electron transport (e.g. Refs. \cite {Falk:2018, Sentoku:2007}), and nonlinear absorption due to high x-ray fluence or short timescale e.g. resulting from mass ejection of core electrons\cite{Vinko:2012}
%[original paper on this, new: https://journals.aps.org/prl/pdf/10.1103/PhysRevLett.119.085001 ?also move up], 
and saturation of absorption\cite{Nagler:2009}, %
% fluorescence, stimulated Raman and Brillouin scattering, second harmonic generation, the Kerr effect or the Pockel effect 
%are not considered.
%[Ralph Pohl - Paper?, REFERENCES].
 can modify initial temperature distributions, %for the finite element simulations, and be included in relevant models, 
 but cooling behavior will be similar.
%
 %
%$Ballistic electron transport in the samples could be relevant in causing a redistribution of thermal energy, e.g. before lattice equilibration occurs, however a
%others?
% this is ; that is,.  %USED[https://journals.aps.org/prl/pdf/10.1103/PhysRevLett.120.025002] 
% suggest hot electron transport and nonlocal energy deposition may play a role on longer timescales and over larger distances, suggesting this issue should be examined experimentally.  
% While local thermal equilibrium is assumed between ions and electrons, thermal equilibrium is not assumed in the target itself: we model the macroscopic temperature evolution out of equilibrium.
%In other words, the incident energy is taken to be converted into heat at equilibrium instantaneously and the calculation proceeds with this assumption.% [
%Ralph Pohl,
%]. Energy is deposited  via excitation of electrons, which diffuse through the material and heat up the ions through electron phonon collisions, in a range of time larger than the duration of the pulse. 
%Here we assume rapid thermalization [
%Ralph Pohl
%] of electrons and ions (i.e. within the first $\sim 10^{-12}$ s), such that we may neglect these effects for our temperature evolution calculations.
With a propagation time across the entire target of $\sim$10$^{-11}$ s, it suffices for our purposes to assume the x-ray beam is incident in all points of the target simultaneously.%propagated instantaneously across the target.%, since conduction processes occur on longer timescales.

% As we consider an isochoric system in our models, cooling occurs via heat transport alone (does this make sense?).]

%\subsubsection{Interface Temperature}

%Initial temperature evolutions of interfaces could be influenced by non-isochoric effects as well, e.g. those due to differential thermal expansion and pressure wave propagation [Ping15 etc.], which are not included in our models and which occur on faster timescales than heat conduction potentially leading to more rapid temperature evolution.  However any interfacial temperature variations in this case would be hydrodynamic and adiabatic in nature.  Thermal conduction alone should change the interface temperatures appreciably only on longer timescales, where substantial heat conduction into surrounding colder regions is possible. 

%\subsubsection{GEOMETRY}
\subsubsection{Materials parameters}\label{Sec:materials} %AND SIMULATION CONDITIONS}

A suite of materials with varying properties are included in the models to examine the possible range of heating and cooling behavior under x-ray irradiation.  As the degree of x-ray absorption in a substance is roughly given as %[\textit{Malcolm: I need some kind of reference for this...}]%[Ref?]
\begin{equation}
\alpha \propto \frac{\rho Z^4}{A E_{photon}^3}
\label{Eq:xra}
\end{equation} 
%where $\rho$ is the density, 
%x-ray energy is $E_{photon}$, 
where atomic number and mass are $Z$ and $A$ respectively, we sought to explore samples over a wide range of $Z$, and lesser variances in the surrounding low-Z materials, as well as a range of photon energy which has a similarly strong effect on absorbance. %, i.e. for a given material of number density $\rho/A$, it is proportional to $Z^4$. 
Material properties are assumed to be constant with temperature, in order to provide a representative and simplified picture of material response for a range of possible materials.  More detailed materials modeling could include temperature (and pressure) sensitivity of parameters, effects of phase transformations, and effects of electronic excitations (e.g. electronic heat capacity\cite{Jeanloz:2007}), for example.  
%and other short timescale (sub picosecond) effects not included in these simulations.  
These models thus provide a representative picture of the lifetime and properties of hot states in strongly-tamped targets following a comparatively rapid emplacement of equilibrium temperature by irradiation.  %Accounting for local excitation and equilibration processes (e.g electron-ion equilibration) on time and length scales significantly smaller than the experimental scale can be included for more detailed prediction of the initial (equilibrium) temperature distribution and sample properties, but the subsequent thermomechanical evolution of target simulated here.t
%Effects of phase change and the scaling of physical parameters with temperature are not included in our simulations.  This includes transient [that PRL] or permanent [Diamond>Graphite papers, 2?] phase transformations, [and nonequilibrium phenomena such as kinetic hindrance[metastablity in shock studies?] of equilibration-rate limited[some FEL melting study] phase transitions [BUT YOU ARENT CONSIDERING ANY PHASE TRANSITIONS?  MIGHT JUST SAY: PHASE TRANSFORMATIONS]. ?]
All material properties are taken to be isotropic.

 %%%%%%%%%%%%%%%%%%%
\begin{table*}
\begin{center}
\centering
\footnotesize\setlength{\tabcolsep}{5pt}
\begin{tabular}{|c|c|c|c|c|c|}
\hline 
\multicolumn{6}{|c|}{Standard Configuration, Finite Element Models} \\\hline
Medium & \multicolumn{3}{|c|}{Materials} & Photon  & {Energy/} \\
 %      Thickness[$\mu$m] & Foil & Medium & Tamper &  Energy[$ke$V] &  Pulse[$\mu$J] \\ \hline \hline   
       Thickness[$\mu$m] & Sample & Medium & Tamper &  Energy[keV] &  Pulse[mJ] \\ \hline \hline 
5 & Fe &  Al$_{2}$O$_{3}$ &  Diamond &  25 &  0.35 \\ \hline \hline
\multicolumn{6}{|c|}{Varying Configurations, Finite Element Models} \\ \hline
%Sec1 &	Sec2	&Sec3             &	Sec4	&Sec5	&Sec6\\\hline
Medium & \multicolumn{3}{|c|}{Materials} & Photon  & Energy/ \\
%       Thickness[$\mu$m] & Foil & Medium & Tamper &  Energy[$ke$V] &  Pulse[$\mu$J] \\ \hline 
       Thickness[$\mu$m] & Sample & Medium & Tamper &  Energy[keV] &  Pulse[mJ] \\ \hline 

0,	&Fe,        &      &Diamond,  &25,	&3.5,\\
2,	&H$_{2}$O,	    &Al$_{2}$O$_{3}$,&Be,	      &20,  &0.35,\\
5,	&Mo,	    &LiF,  &Graphite, &	15, &35,\\
10	&Pb,	    &Ar   &Al$_{2}$O$_{3}$,    &	10, &3.5\\
	&Gd$_{3}$Ga$_{5}$O$_{12}$	&	   &Kapton   &	5   &	\\ \hline
% \multicolumn{6}{l}{A:  Observations: Pure and isotropic materials. }\\
   \end{tabular}
\caption{Model input parameters, with standard configuration at top and sets of varying simulation parameters explored shown at the bottom.}
\label{tab:Experimentalschema}
\end{center}
\centering
\end{table*}
%%%%%%%%%%%%%%%%%%% 
	
The model calculations were performed most commonly with a standard material system comprising a primary sample of iron,  a surrounding medium of alumina (Al$_{2}$O$_{3}$), and diamond as the tamper (Tables \ref{tab:Experimentalschema} and \ref{tab:physicalpropertiesstd}). %and \ref{tab:physicalproperties}
  This standard assembly was then explored by varying independently the x-ray energy (Tables \ref{tab:Experimentalschema} and \ref{tab:physicalpropertiesstd}), beam power (Table \ref{tab:Experimentalschema}), the materials comprising the sample, medium, and tamper (Tables \ref{tab:Experimentalschema} and \ref{tab:physicalproperties}), and the medium layer thicknesses (Table \ref{tab:Experimentalschema}).  Sample materials were chosen to represent a range of possible x-ray absorption levels, including a range of metals across a range of Z (Fe, Mo, Pb), a representative low-Z material (H$_2$O) which is also an insulator, and a representative high-Z insulator (gadolinium gallium garnet, Gd$_3$Ga$_5$O$_{12}$, or `GGG').  %For the properties of Fe we took values representative of Fe at planetary core conditions [Konopkova, Nature, 2016] (?); for others we took their ambient properties.
  The additional material at the outside edge of the sample area, referred to as a gasket, is composed of rhenium (Table \ref{tab:physicalproperties}).   
 Representative thermo-physical and optical bulk material parameters (Tables \ref{tab:physicalpropertiesstd} and \ref{tab:physicalproperties}) were taken from values measured at ambient pressure and temperature, unless otherwise noted.  X-ray photon energies were taken from the hard x-ray regime typically available and used at FEL sources in x-ray diffraction and absorption measurements.  Pulse power (given in terms of total pulse energy) was taken to peak near the maximum presently available at such facilities.
%The input parameters for bulk materials used in the model include  of pure materials, the absorption coefficient and reflectivity showed dependence to wavelengths are listed in %\cite{McWilliams:2015}. 
  
 %%%%%%%%%%%%%%%%%%
\begin{table*}
\begin{center}
\footnotesize\setlength{\tabcolsep}{5pt}
\begin{tabular}{ |l|c|c|c|c|c|c|c|c|}     \hline
\multicolumn{9}{|c|}{Standard Material Parameters, Finite Element Models} \\ \hline
&\multicolumn{3}{|c|}{Thermodynamic Properties}&\multicolumn{5}{|c|}{ Photo absorption coefficient $\alpha$ [1/m]}\\ \hline
 Material & $\rho$ & $C_P$ & $k$ & 25 & 20 & 15 & 10 & 5 \\
  & kg m$^{-3}$ & J (kg K)$^{-1}$ & W (m K)$^{-1}$
  &keV&keV&keV&keV&keV \\ \hline 
Fe & 7870 & 450 & 60 & 1.03$\times 10^4$ & 1.95$\times 10^4$ & 4.40$\times 10^4$	 & 1.33$\times 10^5$ &1.05$\times 10^5$ \\ 
Al$_{2}$O$_{3}$ &3975&765&46& 4.32$\times 10^2$&	8.04$\times 10^2$	&1.86$\times 10^3$&	6.23$\times 10^3$&	4.82$\times 10^4$\\ 
Diamond &3520&630&1500& 9.10$\times 10^1$&	1.28$\times 10^2$& 2.40$\times 10^2$&	7.69$\times 10^2$&	6.68$\times 10^3$\\ \hline
%\multicolumn{6}{l}{*values taken at 300K and atmospheric pressure}\\
\end{tabular}
\caption{Materials parameters used in FE calculations for standard sample configuration.}
\label{tab:physicalpropertiesstd}
\end{center}
\end{table*}
%%%%%%%%%%%%%%%%%%%%%%%%%%%%%%%%%%%%%%%%%%%%%%%%

%%%%%%%%%%%%%%%%%%%%%%%%%%%%%%%%%%%%%%%%
\begin{table*}
\begin{center}
\footnotesize\setlength{\tabcolsep}{5pt}
\begin{tabular}{ | l  |c|c|c|c|p{.5cm}|}     \hline
\multicolumn{5}{|c|}{Additional Material Parameters, Finite Element Models} \\ \hline
%%%%%%%%%%%%%%%%%
%&\multicolumn{3}{|c|}{Material Properties$^{*}$}&\multicolumn{5}{|%c|}{ Photo absorption coefficient $\alpha$ $[1/m]$}\\ \hline
 %Material & $\rho$[$kg$/$m$$^{2}$] & $C_P[$J$/kg K]$ & $k[W/m K]$
  %&$25 keV$&$20 keV$&$15 keV$&$10 keV$&$5 keV$ \\ \hline
  &\multicolumn{3}{|c|}{Thermodynamic }&\multicolumn{1}{|c|}{ Absorption }\\ 
    &\multicolumn{3}{|c|}{Properties}&\multicolumn{1}{|c|}{coefficient (25 keV)}\\ \hline
 Material & $\rho$ & $C_P$ & $k$ & $\alpha$  \\
  & kg m$^{-3}$ & J (kg K)$^{-1}$ & W (m K)$^{-1}$  &[1/m] \\ \hline 
H$_{2}$O &1000&4187&0.686& 4.34$\times 10^1$	\\ 
Mo &10188&251&113& 4.63$\times 10^4$	\\ 
Pb &11340&140&30& 5.28$\times 10^4$	\\ 
Gd$_{3}$Ga$_{5}$O$_{12}$ &7080&381&11& 1.32$\times 10^4$	\\ 
LiF & 2639&1562&11& 1.18$\times 10^2$	\\ 
Ar $^{ a}$ & 5550&570&60& 2.46$\times 10^3$	\\ 
Be  &1848&1825&201& 3.14$\times 10^1$	\\ 
Graphite&2210&830&470 & 5.71$\times 10^1$\\ 
Kapton &1420&1095&0.46& 4.36$\times 10^1$\\ 
Re &21020&140&48&	... $^{b}$\\ \hline
% \multicolumn{5}{l}{$^{*}$:Material properties taken from standard tables.}\\
\multicolumn{5}{l}{$^{a}$Properties taken for high pressure solid Ar, as used in anvil cells\cite{Goncharov:2012}.}\\
 \multicolumn{5}{l}{$^{b}$Value not used in the simulation.}\\
\end{tabular}
\caption{Parameters for other materials used in FE models, including the different materials tested for the sample, medium and tamper, and that used in the gasket.}
\label{tab:physicalproperties}
\end{center}
\end{table*}
%%%%%%%%%%%%%%%%%%%%%%%%%

%\subsubsection{Samples}

%\subsubsection{Tampers}  
 
	Diamond was selected as an ideal tamper due to its high x-ray transparency, high thermal conductivity, and high strength to withstand mechanical stresses generated by heating %and to act as an anvil for applying 
or %application of
%initial stress upon 
pre-compressing samples, as in a diamond anvil cell\cite{Dewaele:1998}. 
%In the context of applying initial stress to samples or withstanding thermal stresses generated during rapid heating, 
Diamond has an extremely high mechanical damage threshold beyond that of all known substances\cite{McWilliams:2010} with ability to withstand localized stresses exceeding a TPa\cite{Dubrovinsky:2012}.   
%Diamond has both high compressive strength (prb) and tensile strength (prl), not only enabling pre compression of samples to high pressures (Sec. X) but resisting thermal expansion and stresses induced by laser heating (LHDACpapers).
It has the highest thermal conductivity of all known bulk matter, allowing it to act as an excellent heat sink which, when properly configured, allows the tamper to remain at very low temperature even when adjacent to very high temperature matter\cite{Dewaele:1998,McWilliams:2015}.  
%In this way its performance is robust in properly designed experiments at extreme conditions.
	Metastable at ambient conditions, and only thermodynamically stable under pressures exceeding $\sim$13 GPa at room temperature, it is generally at risk of damage from thermal decomposition processes such as oxidation and graphitization at temperatures exceeding $\sim$1000 K, as well as non-thermal graphitization at high x-ray fluence\cite{Medvedev:2013}.
	%[] and sublimation [], UV to x-ra radiation-induced graphitisation[2 papers].  
Even under high pressure where diamond is stable, it will melt at sufficiently high temperature\cite{Eggert:2010}.
%	Due to the sensitivity of diamond to temperature (graphitisation, oxidation, sublimation, and melting at higher pressures) we also considered a tamper composed of graphite with representative aggregate properties, that is, relatively high thermal conductivity. It should be noted that should graphite be incorporated into a target as a single crystal, its orientation would be important in defining its performance\cite{Slack:1962}
	%[ Kato01, 
%http://iopscience.iop.org/article/10.1088/0957-0233/12/12/307/pdf, MAYBE ALSO: 
%KavnerRSI08,
%> http://aip.scitation.org/doi/full/10.1063/1.2841173, 
%Rainey2013 
%> http://aip.scitation.org/doi/pdf/10.1063/1.4830274]
%]
Several other plausible tamper materials are considered which can provide qualities including competitive mechanical strength behavior (Al$_2$O$_3$), superior x-ray transparency (Be, Kapton), resistance to thermal degradation and stability over a wide range of temperature (Be, Al$_2$O$_3$, Graphite), and relatively good thermal conductivity within an order of magnitude of that of diamond (Be, Graphite) as well as extremely low thermal conductivity where thermal confinement rather than dissipation may be desired (Kapton).
% Mechanical weakening is also likey to reduce its performance as an anvil should its temperatures approach those required for thermal alteration [Youngbook].  However, crucially, and 

%\subsubsection{Interfacial layer (medium)}

%	The interfacial layer (medium) between the tamper and sample could act as a pressure-distributing medium, either for pre-application of pressure to the sample (as in a DAC) or to distribute thermal stress induced by heating.  This medium can thermally insulate the sample from the tamper, improving its retention of heat and preventing direct damage to the tamper.  For example, for a diamond tamper, this could can prevent local ballistic electron\cite{Medvedev:2013} or thermal\cite{McWilliams:2015} graphitization of the tamper surface.

%\subsubsection{X-ray parameters}

%\subsubsection{\label{sec:generalmethodology} GENERAL SIMULATION METHODOLOGY}

\subsection{Hydrodynamic Models}\label{Sec:ModelsHydro}
%
%\subsubsection{General Context}
%
As the temperature is increased in the targets, hot areas are subject to thermally-driven expansion, and local stresses can develop which are roughly proportional to the amplitude of the temperature change.  On short timescales (fs-ps), heating is fully isochoric, or nearly so. On the longer term (ps-ns), expansion\cite{Levy:2015} and the concomitant production of stress-density waves will occur.  In the limiting case of isochoric heating and assuming hydrostatic stress and LTE conditions, we can consider the thermodynamic identity
\begin{equation}
\left(\frac{\partial P}{\partial T}\right)_V = \beta K_T 
\label{eq:thermalpressure1}
\end{equation}
where $\beta$ and $K_T$ are the volumetric thermal expansivity and isothermal bulk modulus, respectively.  This implies an isochoric thermal pressure change $\Delta P _V$, for a given imposed temperature change $\Delta T$, as 
\begin{equation}
\Delta P _V \simeq \beta K_T \Delta T.
\label{eq:thermalpressure2}
\end{equation}
%Thus thermal pressures on the order of the bulk modulus are achieved at $\Delta T \simeq 1/\beta$.  Taking a typical value of $\beta \simeq 10^{-5}$ K$^{-1}$ for condensed matter, this characteristic temperature would be roughly $10^5$ K, at the upper limit of conditions predicted here.   %Thus, more compressible samples (lower $K_T$) would produce less thermal pressure.  Regardless, 
With $K_T$ of order 1 - 10$^3$ GPa and $\beta \simeq 10^{-5}$ K$^{-1}$ for condensed matter, and considering maximum achieved temperatures in the range of 10$^3$-10$^5$ K,
%10^3$ for condensed matter (considering also a modest increases under pressure) 
thermal stresses produced in typical experiments can reach values between 10$^{-2}$ and 10$^3$ GPa, compatible with the creation of high pressure shock  waves. 

In an unconfined target, the expansion of the heated sample via pressure waves can reduce the amplitude of dynamic stress to zero;
%The corresponding deviation from isochoric conditions will reduce the amplitude of stress, to zero for free expansion (
%ignoring tensile or negative stress states that might occur in high tensile strength materials).  However,
for a tamped target free expansion is prevented leading to a more complex system of compression and release.
%Additionally, while dynamic stresses can relax toward their initial conditions, a well confined sample can retain a degree of thermal stress indefinitely\cite{Dewaele:1998}.
%
%in place until the sample has time to release.  In the system considered here, a crude estimate of this timescale is about twice the time it takes for a sound wave to transit the tamper, or $\sim$ 1 $\mu$s, ignoring the possibility of cracks or, as we will see below, voids formed due to strong tensile stresses. The stress evolution of such a system could be complex, coupling to thermal conduction on similar timescales.
%, and it is beyond the scope of this study to accurately examine this.  
%However, we can outline the nature of these stresses and the manner in which they might affect experiments.
%
%\subsubsection{Methodology}
We have employed the \textsc{hyades} hydrocode\cite{Larsen:1994} to study the 1D evolution of the stress, strain, and temperature in the adiabatic initial part of the experiment following heating.  
Experiments are initialized at $T$=300 K and ambient pressure and density for the different target layers. We use tabular equations of state (Sesame 7830 for diamond, Sesame 2980 for Mo, and Sesame 7410 for Al$_2$O$_3$) in the models.
%[, or optionally, at elevated pressure and density to represent precompression??].  X-ray heating is assumed to be take the form of a pulse of $\sim$100 fs length, which begins at $t=0$.
We model only the first several $\mu$m of the tamper closest to the sample; where wave interactions with simulation cell boundaries produce unphysical conditions, very late in the simulation, the results are removed.
An average atom ionization model is used to generate opacities. LTE is not assumed in these simulations \textit{a priori}, as done for the conduction calculations, due to the faster nature of hydrodynamic phenomena leading to timescales adjoining %/abutting/bordering
 the picosecond regime of equilibration processes. However these equilibration effects occur rapidly only in the first moments of the simulation, and are as expected not relevant to bulk hydrodynamic processes in a target of this size, where shock durations are of order hundreds of picoseconds.
We have not modeled 2D effects which would be important to include to describe accurately the later-time behavior of this system, roughly as wave propagation distances exceed the beam radius.%, or sooner if the beam has a inhomogeneous spot (as included in the finite element models)
%\textbf{Depending on how the sample and simulation is configured, the state produced by these interacting release waves can be one of .}

%We select materials with suitable opacity models to accurately account for the initial heating by the x-ray.

\begin{figure*}%[htbp]
	\centering
	\includegraphics[width=17 cm]{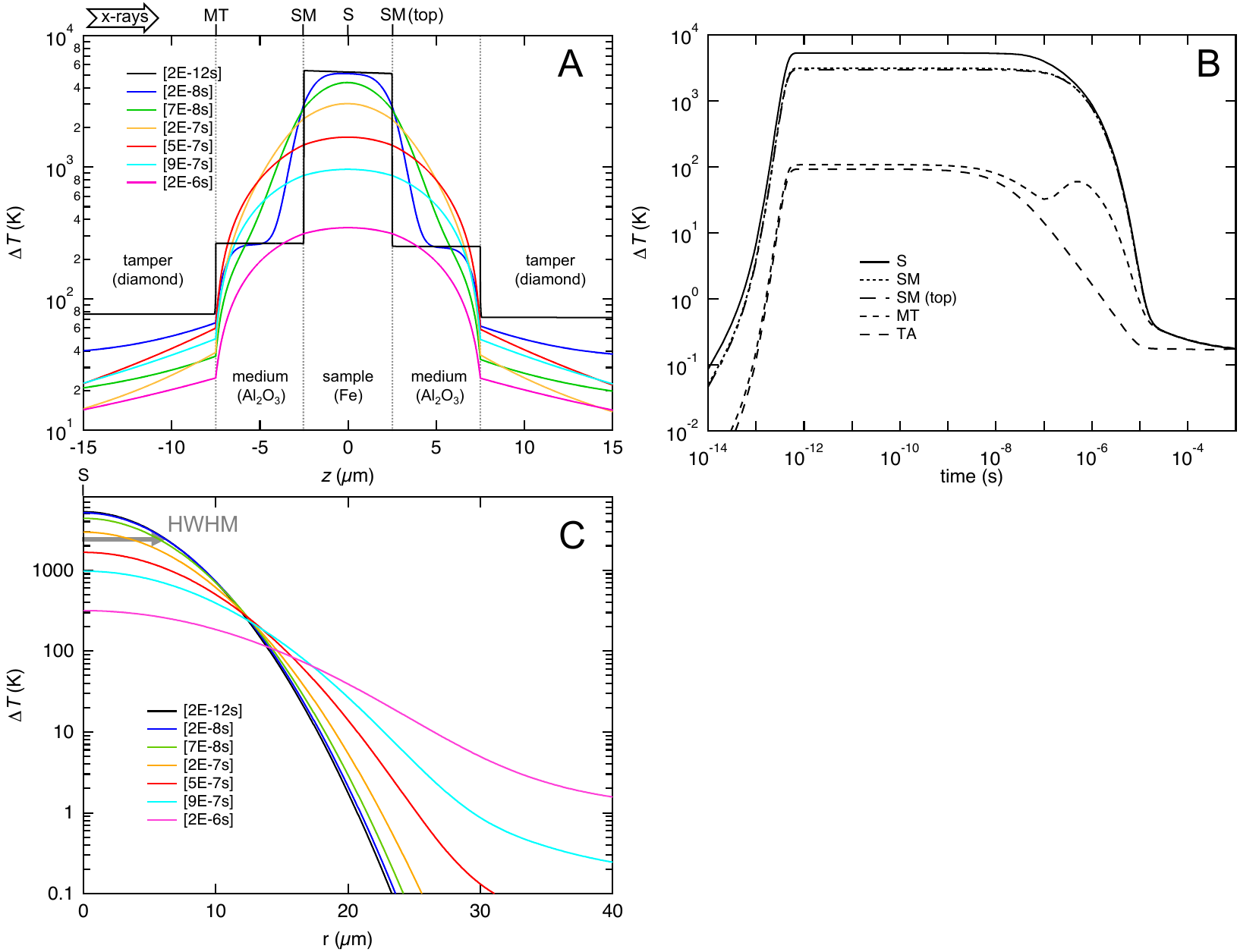} 
	\caption{Thermal response of the baseline simulation. A, Temperature change vs. position along the beam path center ($r=0$) in the sample region. B, Temperature change vs. time at (see Fig. \ref{fig:schemaDAC}) sample center (S), leading (SM) and trailing (SM top) sample-medium interfaces, leading medium-tamper interface (MT), and leading tamper free surface (TA). C, Radial temperature distribution at the center of the sample, showing the half-width at half maximum (HWHM) of the beam and initial temperature distribution (black).  Times are given in square brackets in seconds.}
	\label{fig:ResultsBaseline1}
\end{figure*}

\begin{figure*}%[htbp]
	\centering
	\includegraphics[width=17 cm]{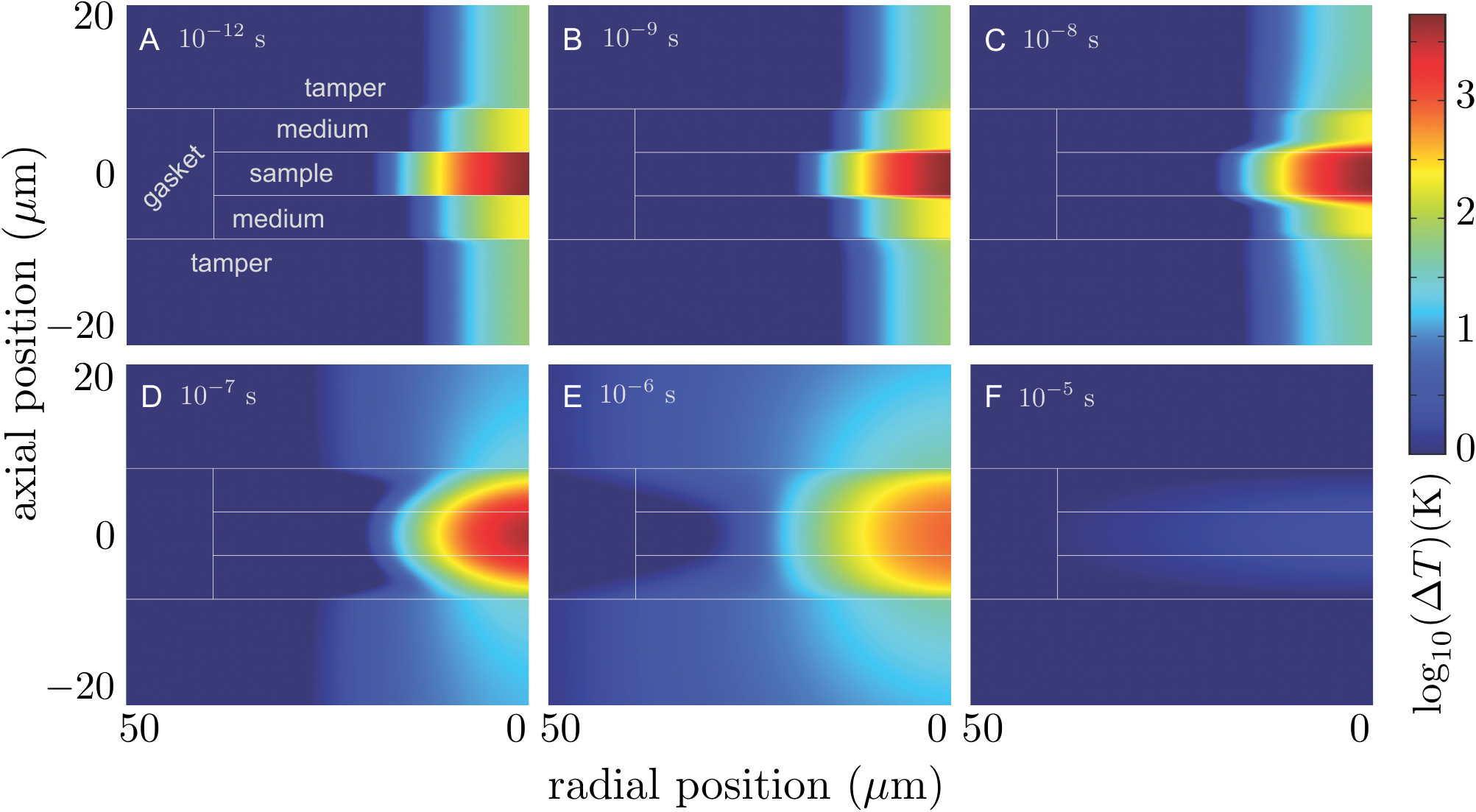} 
	\caption{Temperature change map in the $r-z$ plane for the standard experiment at different times, showing the detailed behavior of the sample area. Lines show the boundaries between sample components (see Fig. \ref{fig:schemaDAC}).} %Thermal equilibrium (i.e. for electronic and ionic temperature) is assumed to be reached within $\sim 10^{-12}$ s. }
	\label{fig:ResultsBaseline2}
\end{figure*}

\section{\label{sec:res} RESULTS}
\subsection{Finite Element Heat Transfer Results}\label{Sec:FEResults}
\subsubsection{Standard configuration}
The baseline simulation, on which other simulations are perturbations, uses the standard target materials arrangement, radiation of 25 keV and a net pulse energy of 0.35 mJ (Fig. \ref{fig:ResultsBaseline1}). 
A close-up view of the sample region (Fig. \ref{fig:ResultsBaseline2}) shows the development of temperature gradients, from an initial state of nearly-constant temperature within layers (at given $r$) and discontinuities at layer interfaces.
The diamond tamper in this case, by virtue of its high thermal conductivity, provides rapid quenching of the tamper itself by radial heat flow, while the sample region remains hot on longer timescales (Fig. \ref{fig:ResultsBaseline2}).
Initial radial gradients (imposed by the assumed Gaussian beam profile) are roughly preserved and somewhat broadened with time (Fig. \ref{fig:ResultsBaseline1}C).  Note the sudden rise in temperature at the medium-tamper interface just before $10^{-6}$ s (Fig. \ref{fig:ResultsBaseline1}B), corresponding to arrival of a heat wave from the sample moving across the medium.

%\begin{figure}[htbp]
%	\centering
%	\includegraphics[width=8.5 cm]{./Figures/3D_sequence_farb.pdf} 
%	\caption{Temperature evolution of the sample area in 3D for the standard experiment showing rapid radial conduction and heat dissipation in the diamond tamper.}
%	\label{fig:TamperEvolution2}
%\end{figure}

\begin{figure*}%[htbp]
	\centering
	\includegraphics[width=17 cm]{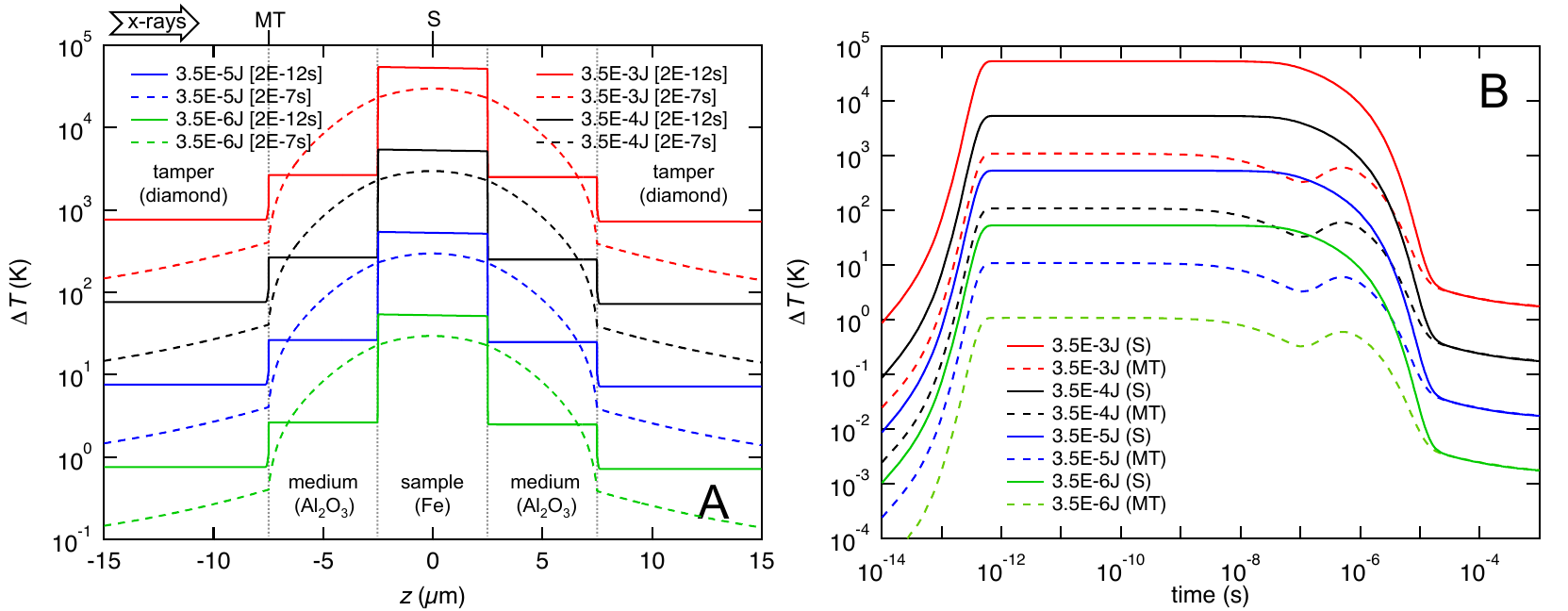} 
	\caption{Variance of thermal response with x-ray fluence (energy per pulse). A, Temperature change vs. position along the beam path center ($r=0$) in the sample region. B, Temperature change vs. time at sample center (S) and leading medium-tamper interface (MT). The black lines correspond to the standard simulation.  Times are given in square brackets in seconds.}§
	\label{fig:ResultIntensity}
\end{figure*}

\subsubsection{Radiation variance: X-ray intensity}

Varying the beam intensity (Fig. \ref{fig:ResultIntensity}) proportionally shifts the thermal response of the target components, a result of the assumed linear absorption process and temperature insensitive material parameters.  Thus, as rule of thumb, the temperature change at any x-ray fluence can be computed from a given simulation's $\Delta T^{sim}$ by scaling to the ratio of the x-ray fluencies, i.e.
\begin{equation}
\Delta T=\frac{I_{max}}{I_{max}^{sim}}\Delta T^{sim}.
\label{Eq:IntensityScaling}
\end{equation}
%to determine the level of heating for different pulse energies $I_{max}$.

\begin{figure*}%[htbp]
%===============================TEMPORARY=====================================
\vspace*{-1cm}
%===============================TEMPORARY=====================================
	\centering
	\includegraphics[width=17 cm]{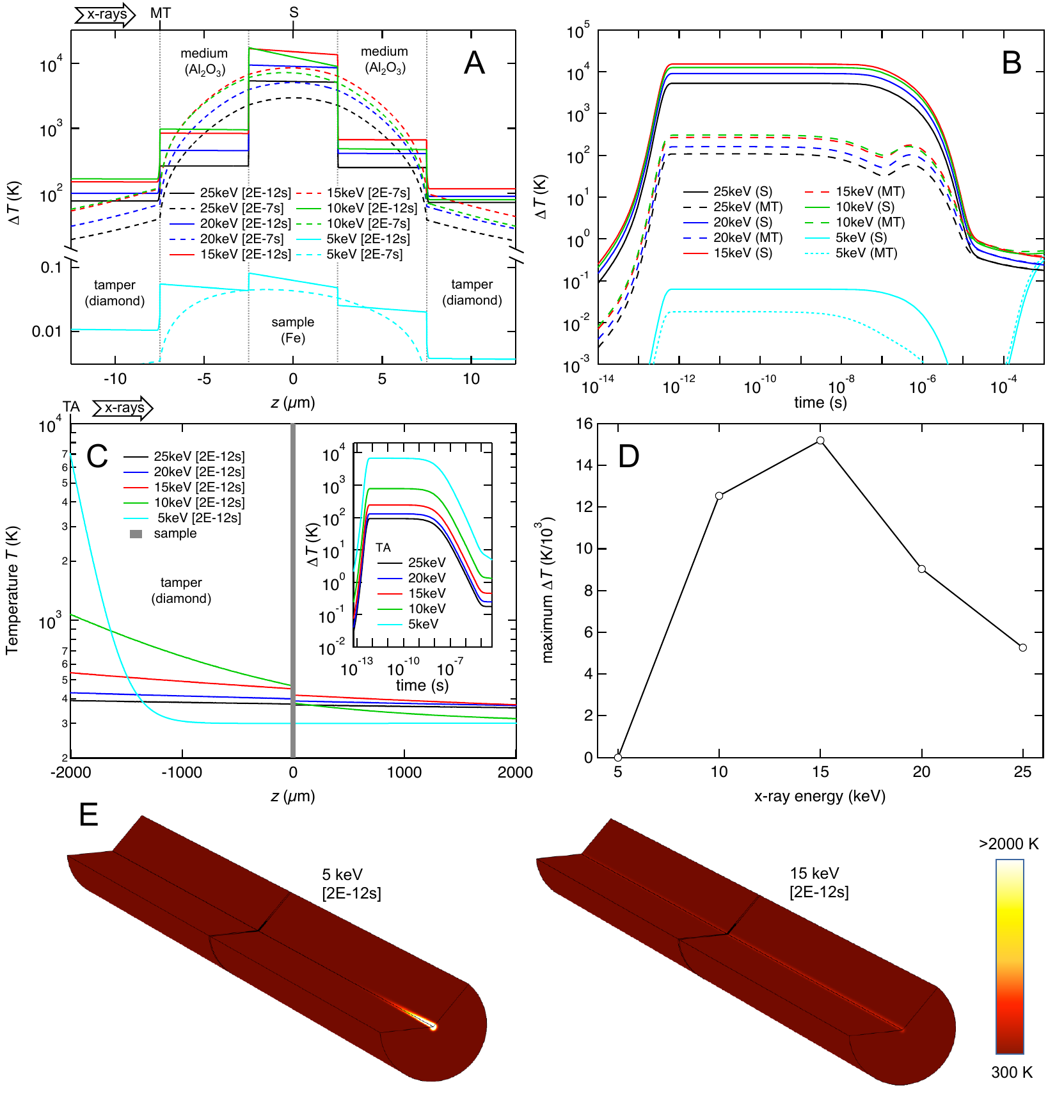} 
	\caption{Variance of the thermal response with x-ray photon energy.  A, Temperature change vs. position along the beam path center ($r=0$) in the sample region.  B, Temperature change vs. time at sample center (S) and leading medium-tamper interface (MT). C, Absolute temperature vs. position along the beam path center ($r=0$) across the whole target, with inset showing temperature change vs. time at the leading tamper free surface (TA). D, Maximum temperature increase at sample center (S) as a function of photon energy.  E, Cylindrical simulation region temperature immediately after heating for 5 keV (left) and 15 keV (right). The black lines in A-C correspond to the standard 25 keV simulation results.  Times are given in square brackets in seconds.}§
	\label{fig:ResultEnergy}
\end{figure*}

\subsubsection{Radiation variance: X-ray photon energy}

Varying the x-ray wavelength (photon energy) through the hard x-ray range will vary the differential absorption in samples, and the temperature gradients established (Fig. \ref{fig:ResultEnergy}).  For lower energies ($\sim$5 keV) the x-ray is absorbed almost entirely within the leading tamper layer (Fig. ~\ref{fig:ResultEnergy}C,E) whereas harder x-rays ($\sim$25 keV) will largely pass through the sample assembly without generating much heating.  
Homogeneity of heating depends on the x-ray energy, with harder x-rays producing superior initial homogeneity %in the adiabatic interval while pronounced initial asymmetries in temperature in the sample area occur for 
and lower energies greater initial asymmetry (Fig. ~\ref{fig:ResultEnergy}A). 
In terms of providing an optimum heating solution, a 15 keV energy provides maximum sample heating, nearly homogeneous temperature in the sample and moderate but survivable heating in the tamper.   

\begin{figure*}%[htbp]
	\centering
	\vspace{-1cm}
	\includegraphics[width=17 cm]{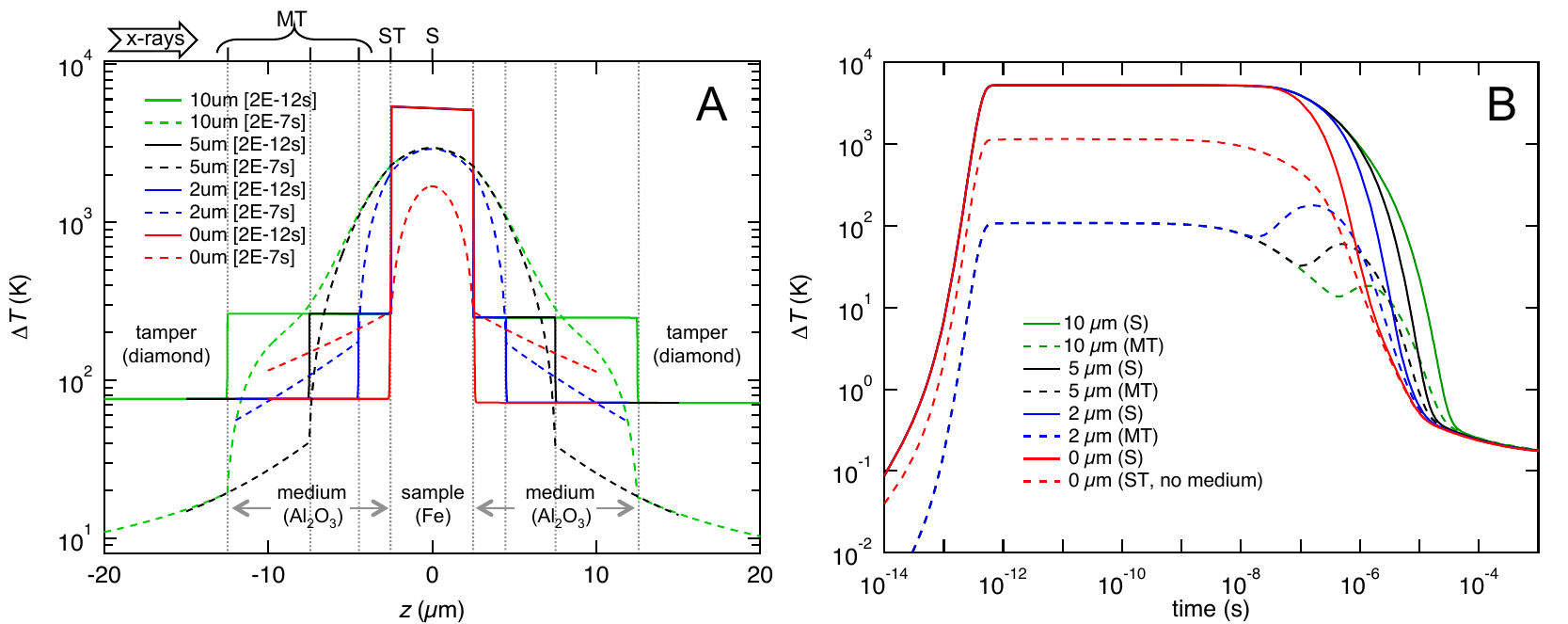} 
	\caption{Variance of thermal response with interfacial layer (medium) thickness.  No layer (direct contact of sample and tamper) corresponds to red.  A, Temperature change vs. position along the beam path center ($r=0$) in the sample region.  B, Temperature change vs. time at sample center (S) and at the leading medium-tamper interface (MT), or sample-tamper interface (ST) in the absence of a medium layer. The black lines correspond to the standard simulation.  Times are given in square brackets in seconds.}§
	\label{fig:ResultThickness}
\end{figure*}

\subsubsection{Geometry variance: Medium thickness}

Without an interfacial medium layer between the sample and tamper, the temperature of the tamper is maximized by direct exposure to the hot sample; the sample is also cooled rapidly, but the tamper interface remains relatively hot (Fig. \ref{fig:ResultThickness}). Addition of even a thin medium layer reduces the temperature in the tamper considerably, while slowing sample cooling.  When a medium is present, sample cooling behavior is insensitive to medium layer thickness, up to $10^{-7}-10^{-6}$ s, after which it varies considerably. 
%NOT TRUE (though it depends more on its composition, as discussed in Sec. \ref{sec:MediumVar}). 
Tamper cooling also proceeds more rapidly for a thicker medium layer. Arrival of the heat wave from the sample (Fig. \ref{fig:ResultThickness}B at $\sim$$10^{-6}$ s) can briefly drive tamper interfacial temperatures higher, possibly to above the initial temperature, though this temperature excursion remains below that which would occur in the absence of the medium.  Thus, addition of even a thin medium layer can reduce heating of the tamper and potentially improve its stability.

\begin{figure*}%[htbp]
	\centering
	\includegraphics[width=17 cm]{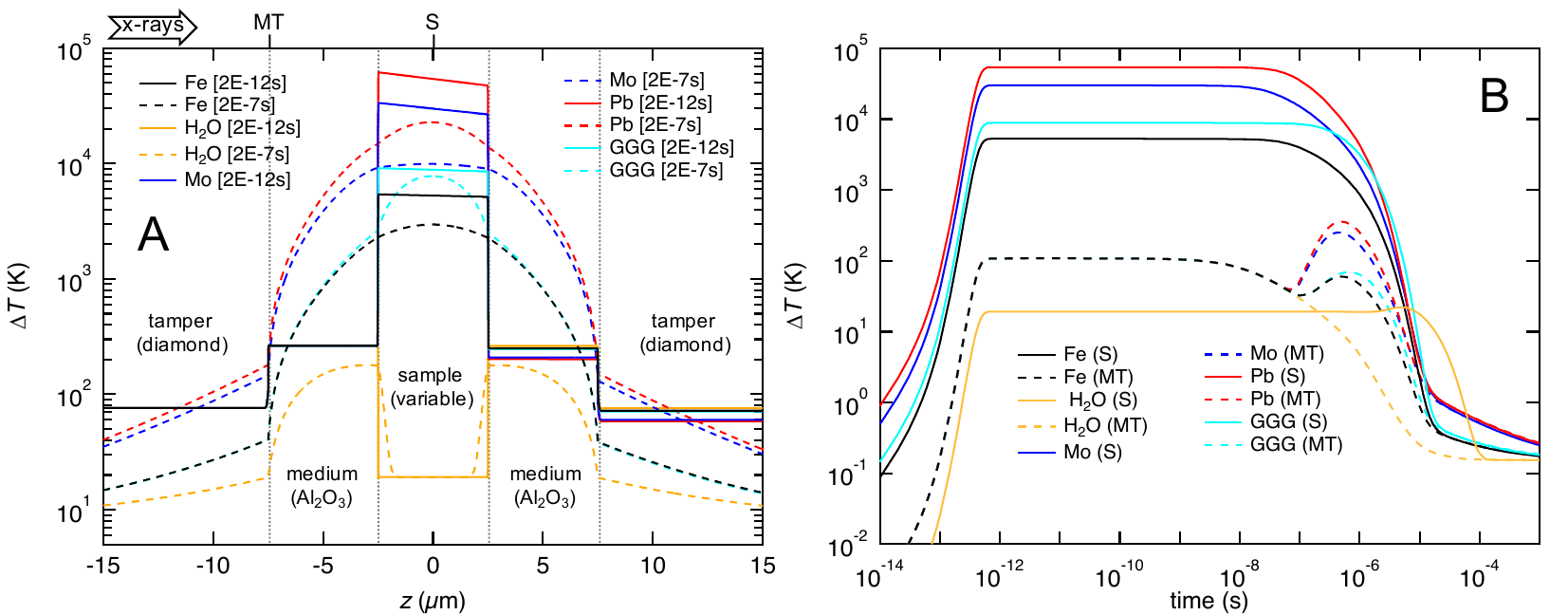} 
	\caption{Variance of thermal response with sample material.  A, Temperature change vs. position along the beam path center ($r=0$) in the sample region.  B, Temperature change vs. time at sample center (S) and leading medium-tamper interface (MT). The black lines correspond to the standard simulation.  Times are given in square brackets in seconds.}§
	\label{fig:ResultSample}
\end{figure*}

\subsubsection{Material variance: Sample}

The samples were generally selected (Fig. \ref{fig:ResultSample}) to exhibit the strongest heating of all target components, and are hence higher-Z materials, with the exception of water which has exceptionally weak heating, below all the other target components.
Electrically insulating samples H$_2$O and the heavy oxide Gd$_{3}$Ga$_{5}$O$_{12}$ (which heats similar to Fe) have reduced thermal conductivities compared to the metals Fe, Mo, and Pb (Table \ref{tab:physicalproperties}), which slow their thermal evolution during the experiments, effectively maintaining the sample temperature even while metals cool off (Fig. \ref{fig:ResultSample}).  Heat waves incident on the tamper, at around $10^{-6}$ s, cause large jumps in tamper surface temperature to well in excess of its initial temperature for hotter samples (Fig. \ref{fig:ResultSample}B).  For water, heat conducts into the sample from the hotter medium layers, leading to a late increase in temperature for this sample.  At this x-ray energy (25 keV) the absorbance of each material is small such that the downstream temperatures are only weakly affected by the different samples (right side of Fig. \ref{fig:ResultSample}A).  Initial asymmetries in temperature in the sample area are more pronounced for the higher Z samples (Fig. \ref{fig:ResultSample}A). 

\begin{figure*}%[htbp]
	\centering
	\includegraphics[width=17 cm]{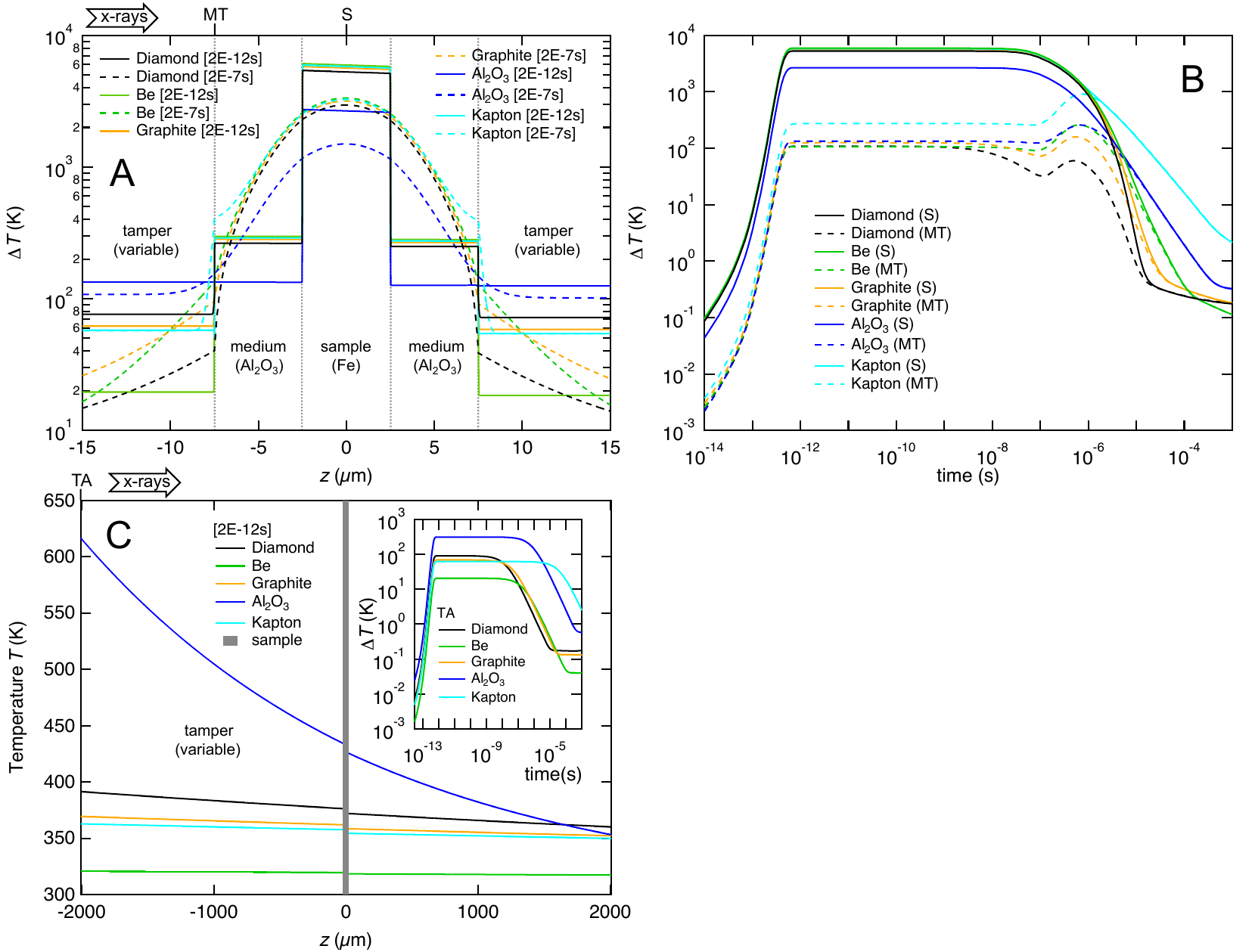} 
	\caption{Variance of thermal response with tamper material.  A, Temperature change vs. position along the beam path center ($r=0$) in the sample region.  B, Temperature change vs. time at sample center (S) and leading medium-tamper interface (MT). C, Absolute temperature vs. position along the beam path center ($r=0$) across whole target, with inset showing temperature change vs. time at the leading tamper free surface (TA).  The black lines correspond to the standard simulation. Times are given in square brackets in seconds.}§
	\label{fig:ResultTamper}
\end{figure*}

\subsubsection{Material variance: Tamper}

The tampers chosen for modeling (Fig. \ref{fig:ResultTamper}) generally show comparable x-ray transparency, with the exception of Al$_2$O$_3$ which has somewhat reduced transmission and hence results in lower sample temperature and higher tamper body temperatures. % due to direct x-ray absorption.
There is significant variance in the temperature and its evolution in the tamper bodies (Fig. \ref{fig:ResultTamper}C inset), but on shorter timescales sample conditions do not evolve differently for the different tampers (Fig. \ref{fig:ResultTamper}A-B). Significant differences in sample temperature evolution are observed only on long ($>10^{-6}$ s) timescales (Fig. \ref{fig:ResultTamper}B).  For the comparably low thermal conductivity plastic (Kapton) tamper, an accumulation of heat at the tamper interface is observed (Fig. \ref{fig:ResultTamper}A), %though with limited effect on sample temperature evolution compared to high thermal conductivity tampers, up to $\sim$ 1 $\mu$s.  The heat accumulation at low thermal conductivity tamper interfaces 
which could promote tamper damage.

\begin{figure*}%[htbp]
	\centering
	\includegraphics[width=17 cm]{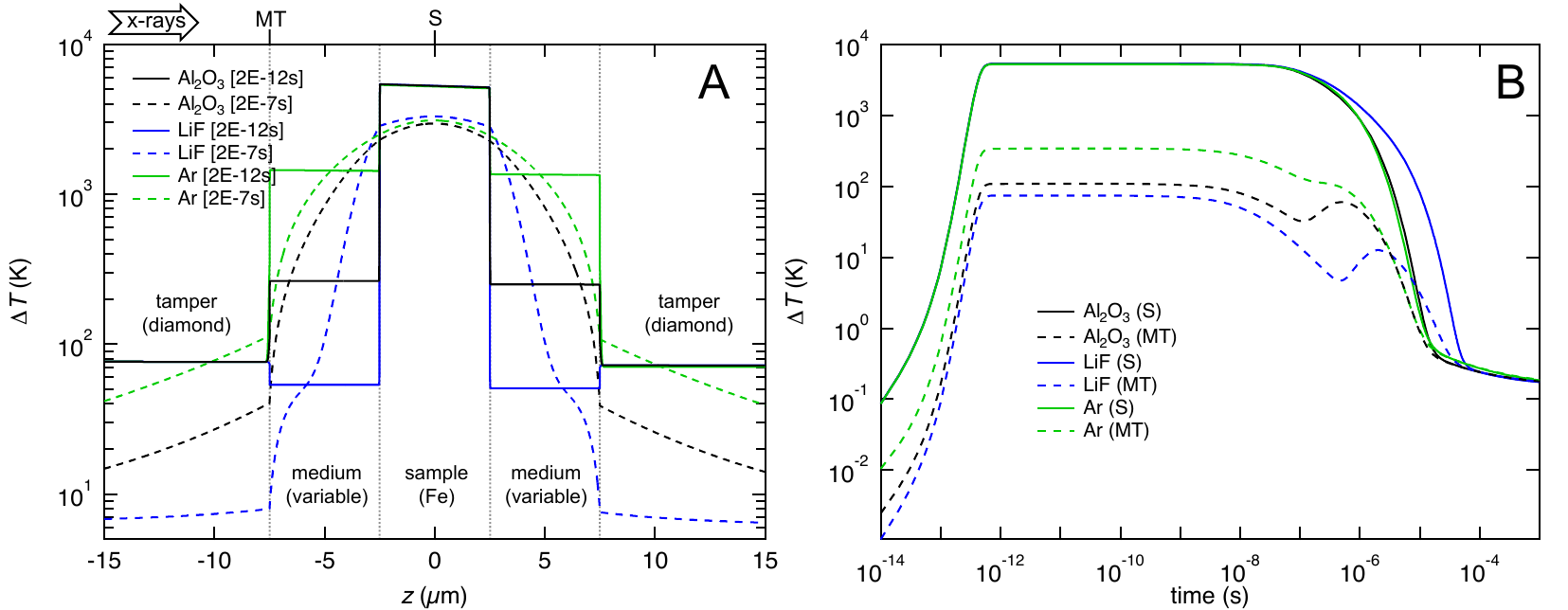} 
	\caption{Variance of thermal response with interfacial layer (medium) material.  A, Temperature change vs. position along the beam path center ($r=0$) in the sample region.  B, Temperature change vs. time at sample center (S) and leading medium-tamper interface (MT). The black lines correspond to the standard simulation. Times are given in square brackets in seconds.}§
	\label{fig:ResultMedium}
\end{figure*}

\subsubsection{Material variance: Medium}\label{sec:MediumVar}

The interfacial medium layer material selected (Fig. \ref{fig:ResultMedium}) influences the sample temperature by controlling the rate of sample cooling, which is most notable on longer ($>10^{-6}$ s) timescales.  As all media chosen are of low x-ray absorbance, differences in performance are due mainly to the thermal conduction properties of the medium layers.  Sample cooling is most sluggish for the lowest thermal conductivity medium (LiF), even though the initial temperature of this layer is also the lowest (which promotes more rapid cooling, all else being equal).

\subsubsection{General features of target thermal evolution}\label{sec:GeneralObservations}
  
Excluding the heat deposited by the x-ray irradiation, targets of the length scales described are effectively adiabatic on timescales up to 1-10 ns.
As a consequence, considering irradiation on the timescales of typical FEL (10-100 fs) or synchrotron bunch (10-100 ps) sources, there should be little difference between peak temperature and subsequent thermal evolution, once LTE is achieved.  Differences will appear only in the heating rate and potentially arise from nonlinear and ultrafast phenomena sensitive to this rate, but broadly, pulsed x-ray heating in the fs-ns range (Table \ref{tab:facilities}) will produce essentially similar target responses, since these timescales do not allow significant cooling during the energy deposition phase.    Thus for fast sources, the principal parameter for assessing the temperature following x-ray illumination is the total pulse energy and its spatial distribution.  Therefore the thermal evolution calculations made here are relevant for pulses of any length, up to the adiabatic limit of $\sim$ 10 ns.
%This changes for much thinner targets\cite{Ping:2015}, since adiabaticity depends on length and time scales.
 
 In these simulations
%, under isochoric conditions, 
interface temperatures between differentially heated surfaces are effectively constant on shorter (adiabatic) timescales.  %This is because, 
Immediately after heating, the interface achieves a temperature intermediate to that in the bulk of the contacting layers, defined in part by the bulk temperatures and in part by the layer thermal transport.  
These results are confirmed by the analytical solution for interfacial temperatures following rapid emplacement %formation/imposition/
of an interfacial temperature discontinuity\cite{Grover:1974,McQueen:1990}.
%Turcotte and Shubert1982NewGETTHISCHECK, SchmittJGR1989,YooPRL93refsGroverUrtiewJAP1974&McQueenJGR1990
For assumed constant layer thermal conductivities (Sec. \ref{Sec:materials}%\ref{tab:physicalproperties}
), the interface temperature $T_i$ is given as 
%  Indeed interface temperatures (due to the assumed constant thermal conductivity) can be written, assuming constant thermal conductivity?, as X (Yoo has expressions and refs), and so for the cases above 
\begin{equation}
T_i=T_A+(T_B-T_A)/(1+\sqrt{{\kappa_A}/{\kappa_B}}) \label{eq:interfaceT}
\end{equation}
%where $\alpha$ is given by the ratio of thermal diffusivities as
%\begin{equation}
%\alpha=\left(\frac{k_1 \rho_2 C_{P2}}{k_2 \rho_1 C_{P1}}\right)^\frac{1}{2}
%$\alpha^2=$ %^{\frac{1}{2}}%=\left(\frac{k_1}{\rho_1 C_{P1}} \frac{\rho_2 C_{P2}}{k_2}\right)^\frac{1}{2}
%\end{equation}
where subscripts indicate the contacting layers A and B.  This closely predicts the simulated constant interface temperatures before cooling begins (after $\sim$10$^{-8}$ s); e.g. in the baseline model at the leading interface between sample and medium, Eq. \ref{eq:interfaceT} predicts %for the thermal diffusivity and simulated temperature differentials 
an initial interface temperature of $\sim 3200$ K, compatible with the modeled value (Fig. \ref{fig:ResultsBaseline1}) of 3400 K.
 
For targets involving an additional low-Z (medium) layer between the sample and the tamper, a late rise in tamper temperature occurs as the heat wave from the high-Z sample reaches the tamper surface.  The associated heating is often relatively minor, even where extreme sample temperatures are reached: e.g. for $\sim$55,000 K in a Pb sample (Fig. \ref{fig:ResultSample}) the heat pulse only raises the temperature at the tamper surface from $\sim$400 to $\sim$650 K. 
%This heat pulse is equivalent to that described for a two-layer differentially heated target in an XFEL\cite{Ping:2015}. 
The timing and amplitude of the heat pulse is correlated with many properties of the system, showing, for example, a direct correlation with the thermal conduction properties of the materials. It can be observed that the arrival time of this pulse increases systematically with thermal diffusivity of the medium (Fig. \ref{fig:ResultMedium} and Tables \ref{tab:physicalpropertiesstd} and \ref{tab:physicalproperties}), i.e. it is fastest for a layer of dense argon ($\kappa=1.9\times10^{-5}$ m$^2$/s), slowest for LiF ($\kappa=2.7 \times 10^{-6}$ m$^2$/s), and intermediate for alumina ($\kappa=1.5 \times10^{-5}$ m$^2$/s). %, suggesting a potential for thermal transport measurements \cite{Ping:2015, Konopkova:2016,Goncharov:2012}. 
The pulse amplitude is lowest for higher thermal conductivity tampers and highest for the insulating tamper (Fig. \ref{fig:ResultTamper}).  
% by the initial magnitude of differential heating (Fig. \ref{fig:resultssample}), but crucially, also by the tamper properties (Fig. \ref{resultstamper}); i.e. it is lowest for higher thermal conductivity tampers and highest for the insulating tamper.

Comparison of the temperature at the sample center and near the interface between the sample and its surroundings provides some indication of the temperature gradient occurring in the sample. On shorter timescales the temperature distribution in the sample is defined exclusively by the absorption profile (Fig. \ref{fig:ResultEnergy}) with an asymmetric gradient in initial temperature along the beampath (axial direction) possible in low keV experiments (Fig. \ref{fig:ResultEnergy}) or when using high-Z samples (Fig. \ref{fig:ResultSample}).  With time, the sample temperature becomes more symmetric in the axial direction, regardless of the initial heating symmetry, with the lowest values near interfaces and the center remaining warmer.  

For harder x-rays (15 keV and above), peak temperatures in the low-Z tamper are generally produced adjacent to the sample layers, either immediately upon heating (due to interfacing with a hotter medium (Fig. \ref{fig:ResultMedium}) or sample (Fig. \ref{fig:ResultThickness}) layer, or after the heat wave from the cooling sample reaches the tamper (Figs. \ref{fig:ResultThickness}B, \ref{fig:ResultSample}B, \ref{fig:ResultTamper}B).  At lower keV, the hottest portion of the tamper is the leading free surface due to efficient absorption of the beam, however only at the lowest x-ray energy simulated (5 keV) is the tamper hotter than the sample (indeed, there is negligible heating in the sample in this instance).

% \subsubsection{Interface temperatures}

\begin{figure}%[htbp]
	\centering
	\includegraphics[width=8.5 cm]{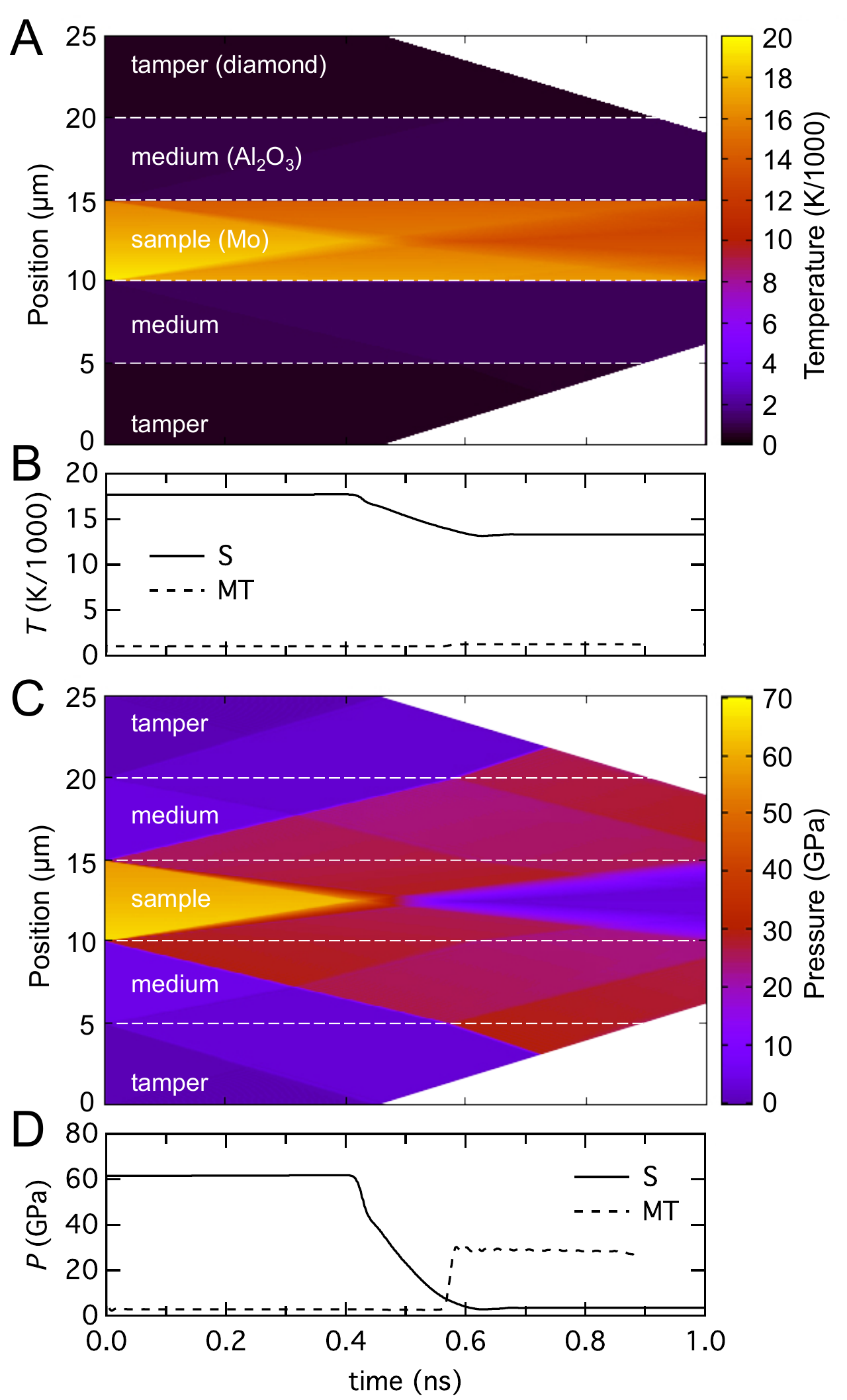} 
	\caption{One dimensional radiation hydrocode (\textsc{hyades}) model for the sample area of a target in first $10^{-9}$ s after irradiation. Here a Mo sample (5 $\mu$m), surrounded by Al$_2$O$_3$ medium layers (5 $\mu$m) and diamond tampers (with thickness truncated to the displayed 5 $\mu$m), is irradiated in a vacuum by 25 keV x-rays (see Fig. \ref{fig:ResultSample}, dark blue curves, for a finite element model of a comparable system, at a different initial temperature).  X-rays are incident from below.  A, Temperature throughout the simulated region (as a function of Lagrangian position and time). B, temperature histories at the sample center (S) and medium-tamper interface (MT).  Temperature changes are adiabatic in nature on this timescale.  C, Pressures throughout the simulated region and time domain. D, pressure histories at the sample center and tamper surface. 	%A,C, pressure evolution and B,D, temperature evolution for a Mo (A,B) and Fe (C,D) sample.
	%The sample is X, 
	%The medium is dense Ne, and the tamper is diamond.  Sample and medium layers are 3 $\mu$m.
	%Only the nearest are of the tamper is included in the simulation. Experiments begin at vacuum and room temperature.  The power level is X; the EOS's used are X (Diamond), X(Ne), X(Fe?), X(Mo).
	Regions where the simulation boundary interfered with the results were removed.
	%Artifacts (releases) appear when the waves strike the arificial simulation boundary within the tamper in some simulations.
	}§
	\label{fig:HYADESambient}
\end{figure}

\subsection{Hydrodynamic Model Results}\label{Sec:ResultsHydro}

%(discuss earlier in detail tension, spall, vapour effects and manifestation in \textsc{hyades})

A representative hydrodynamic model of the initial thermomechanical evolution of a target after irradiation is shown in Fig. \ref{fig:HYADESambient}. Here a Mo sample, contained by an alumina medium and diamond tamper (c.f. Fig. \ref{fig:ResultSample}), is heated with 25 keV x-rays 
at $\sim$$10^{15}$ W/cm$^2$ %checks out after changes.
for $\sim$100 fs to peak temperature near $2\times10^4$ K. 
%\textbf{The initial temperature conditions are comparable to the finite element result for Mo in Fig. \ref{fig:ResultSample}}.  
%The initial temperature conditions are somewhat less than those in the finite element model of the equivalent Mo sample arrangement (Fig. \ref{fig:ResultSample}) due to lower irradiance in the present model.

\begin{figure}%[htbp]
	\centering
	\includegraphics[width=8.5 cm]{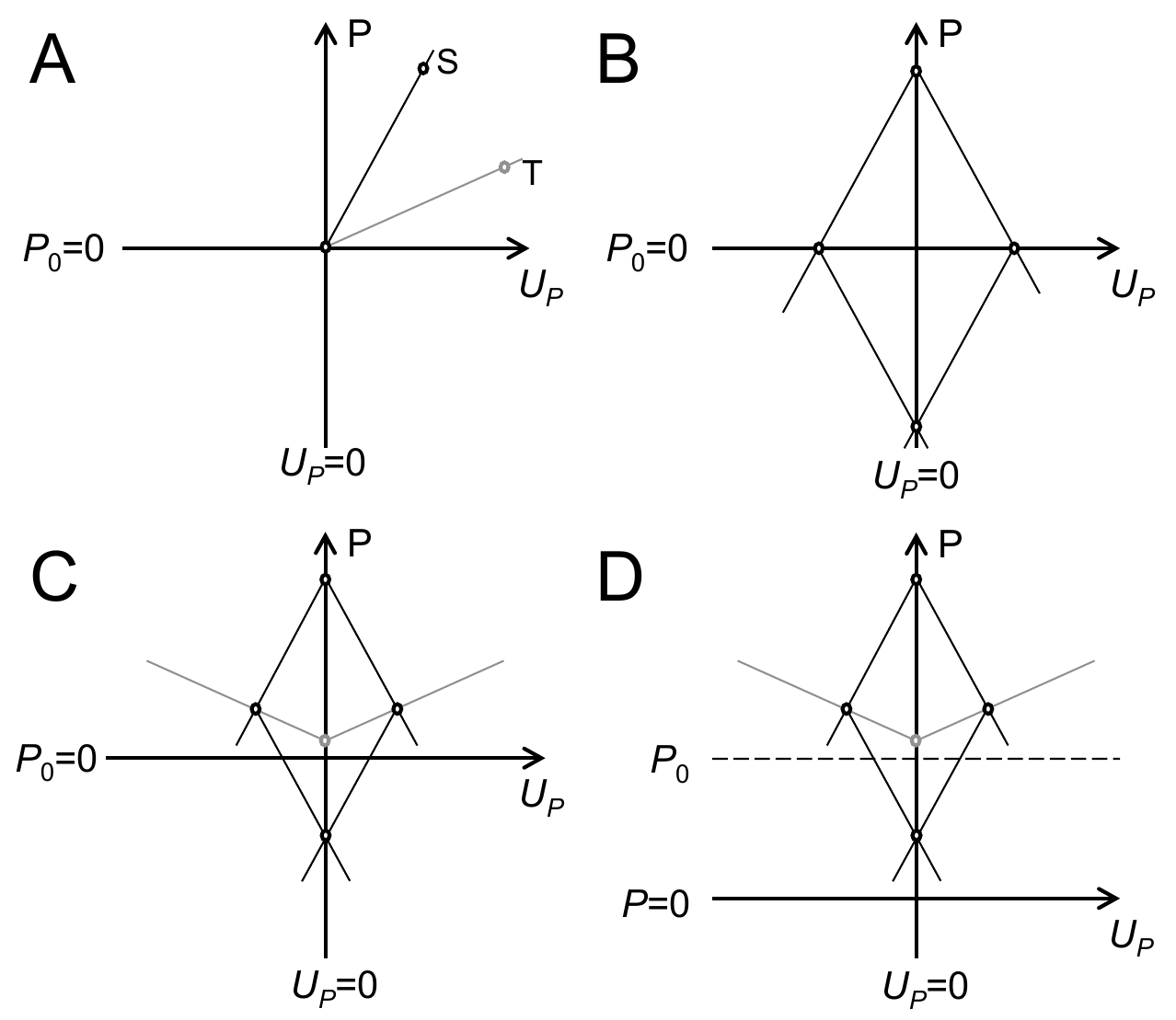} 
	\caption{Impedance match construction for the mechanical evolution of the x-ray heated sample (pressure $P$ vs. mass velocity $U_P$).  Material responses are lines, whereas dots are specific states achieved; S represents the sample and T a surrounding (i.e. tamper) material, presumed to be more weakly heated.  Shocks and releases are approximated as linear elastic (i.e. $\Delta P \approx \rho c_S \Delta U_P$ where $\rho$ is density and $c_S$ is a wave velocity%, such that $P \propto U_P$
	).  Uniform heating in each layer is assumed.
	A, Compression and release response of the high-Z sample (S) and a low-Z tamper (T), where the tamper is assumed to also have lower impedance. Lines indicate achievable states on compression from initial state $P_0=0$, $U_P=0$; the dots represent particular compressed states.
	B, Case of a freestanding sample layer in vacuum under x-ray heating. The sample foil is immediately driven to a high thermal pressure at zero velocity, and releases from both sides (Fig. \ref{fig:HYADESambient}), driving each side of the target to plus or minus a particle speed and zero pressure.   These release waves converge at target center, causing a further stress reduction equivalent to the initial thermal pressure; i.e. the interacting release waves produce tension, and, if it exceeds of the tensile strength of the material, spall. 
	C, Case of a tamped sample, with only a partial reduction in pressure on initial release due to confinement by surrounding material (Fig. \ref{fig:HYADESambient}), and a reduced but not eliminated tension state (tension is prevented if sample and tamper have closer impedances).
	D, While the preceding scenarios A-C apply for a typical laboratory condition with an initial pressure $P_0$ much less than the dynamic pressure (i.e. vacuum or ambient initial conditions), this scenario begins at a high initial hydrostatic pressure ($P_0>0$) comparable in magnitude to the dynamic pressure, as is made possible by pre-compression with a strong tamper\cite{Jeanloz:2007,Armstrong:2010}.  Achieved pressures are larger, while tension is suppressed.}§
	\label{fig:IM}
\end{figure}

Coincident with the heating, the sample layer experiences an increase in pressure to 55-70 GPa, whereas minor heating in the surrounding layers produces weaker initial pressurization. Due to the differential heating and resulting differential pressures, waves of compression or release emerge from interfaces between the heated layers\cite{Sentoku:2007}.  In this hydrodynamic model, the hot, and hence high pressure sample layer undergoes release of pressure as it expands and compresses the cold surrounding layers, driving them to higher pressure.  The sample expands beginning at its surfaces via an inward-moving release wave, while shock waves are driven outward through the medium and toward the tamper.  While this initial process reduces the pressure in the sample, it is not to zero due to the presence of the medium and the requirement of impedance matching at the sample-medium interface (Fig. \ref{fig:IM}C).  This also requires the corresponding shock pressure to be some fraction of the initial thermal pressurization.

The outward moving shocks reflect off the tampers and back toward the sample (at $\sim$0.6 ns), producing a stress maximum on the tamper comparable in magnitude to the initial thermal stress induced in the sample (Fig. \ref{fig:HYADESambient}D).
%The corresponding uniaxial compressive stresses encountered in the tamper are sufficiently low to be resisted by many tamper materials, including diamond. 
A more compressible medium reduces this initial shock stress at the tamper for similar initial sample conditions.  
%Consideration of 2D effects could reduce the amplitude of the pressure wave at the tamper further, due to lateral release effects.
%.  state of tensile stress which in our simulation is supposed to immediately spall the foil (i.e. the target is taken to have no tensile strength).  Beyond
% this point a void appears at the sample center and grows in size as the new foil interfaces with the void  [may] vaporize [, manifesting as a substantially cooled region of the interfacial material].  The internal tension and spall effect is similar to that occurring in volumetric heating of an unconfined foil\cite{Ivanov:2003}, in that it can be thought of as resulting from the large imposed pressure differential between the foil and its surroundings.
%(with the interior foil separating, this process takes a form similar to a flyer-plate like effect).  
%[Ultimately the dynamic pressure is released by the void.]
%While there is no obvious way for the void to collapse in a 1D model, its possible it would eventually due to the localized nature of heating, where shear forces with laterally adjacent cold target would ultimately provide the restoring force to close the void, for a sufficiently thick and strong tamper.  
%[Cavitation and jetting effects are possible if the voids, now bearing vapor, are closed, leading to particularly extreme states in this central region.]
Meanwhile, the inward moving release waves in the sample layer interact in the target center, producing (beyond $\sim$0.5 ns) a stress minimum in the sample which essentially restores the initial (zero) pressure condition. 
These colliding release waves can also produce tensile stress in the target\cite{Ivanov:2003} (Fig. \ref{fig:IM}C-D), which was seen in separate \textsc{hyades} simulations if using suitable mechanical equations of state for the sample layer, and keeping peak stress sufficiently low.
%While heating, 
Compression and release is nearly symmetric about the sample center in Fig. \ref{fig:HYADESambient}, due to near-homogenous heating of each layer at 25 keV; strong asymmetry occurs for inhomogeneous heating in other simulations (e.g. if lower x-ray energy is used).

As confirmed by the simulations, hydrodynamic timescales relevant to a target of this thickness are sufficiently large that LTE conditions can be assumed in the materials.   The hydrodynamic processes are thus comparable to those in conventional shock experiments of durations of order nanoseconds, such as those produced by optical laser pulses\cite{McWilliams:2010,Eggert:2010,Falk:2018,ng:2005,Kraus:2017,Armstrong:2010}.  In such experiments, assuming conditions of thermodynamic equilibrium (i.e. in which materials follow an equilibrium equation of state) is a reasonable approximation.  Simple thermodynamic calculations can thus predict essential details of the hydrodynamics, and find results consistent with the numerical models. %Sec \ref{Sec:DiscussionShock}).
For example, the magnitude of initial pressure can be reasonably predicted as an isochoric thermal pressure, after Eq. \ref{eq:thermalpressure2}.  For the 17700 K temperature rise in the Mo foil, having $K_T=268$ GPa and $\beta=1.50 \times 10^{-5}$ K$^{-1}$, Eq. \ref{eq:thermalpressure2} gives $\Delta P _V \simeq$ 70 GPa; this compares well with the $\sim$62 GPa initial pressure rise calculated using \textsc{hyades} (Fig. \ref{fig:HYADESambient}).

Dynamic stresses should largely relax in $\sim$$10^{-9}$ s, before heat conduction initiates but with permanent and potentially significant effects on the temperature distribution in the target.
%Conditions are adiabatic during the hydrodynamic relaxation, 
Both shock (adiabatic) and release (isentropic) processes modify temperatures (Fig \ref{fig:HYADESambient}A-B). The temperature in the medium and tamper are somewhat increased by shock, however more pronounced is the temperature reduction in the sample during its release.  This expansion cooling can be described accurately with a thermodynamic model, 
% an effect considered analytically in Sec. \ref{Sec:DiscussionShock}. 
taking an isentropic expansion (entropy $S$ constant) of the Gr\"uneisen form 
\begin{equation}
\gamma=-\left(\frac{\partial \ln T}{\partial \ln V}\right)_S,
\end{equation}
where $V$ is the specific volume. The Gr\"uneisen parameter
\begin{equation}
\gamma=\frac{\beta K_T V}{C_V},
\end{equation}
where $C_V$ is the specific heat capacity at constant volume, is often found to follow the relationship 
\begin{equation}
\gamma=\gamma_0 \left(\frac{V}{V_0}\right)^q
\end{equation}
where the subscript `0' indicates reference (here ambient) conditions and the exponent $q$ is of order 1.  Taking starting conditions of temperature and volume as $T_0$ and $V_0$, initial isochorically-heated equilibrium conditions $T_1$ and $V_1=V_0$, and hydrodynamically-released conditions $T_2$ and $V_2$, and assuming constant thermal expansivity and complete release of thermal pressure, %to ambient pressure, 
we have
\begin{equation}
V_2=\left[\beta(T_2-T_0)+1\right]V_0,
\end{equation}
%if thermal expansivity is constant, %and phase changes are ignored,
i.e. the volume of the expanded state $V_2$ is equivalent to that produced on isobaric heating to the same temperature.  Taking $q=1$ we obtain
\begin{equation}
T_2=T_1 \exp \left[ -\gamma_0 \beta (T_2-T_0) \right].
\end{equation}
Solving for an initial temperature $T_1=17700$ K in Mo, with %$\beta=1.50 \times 10^{-5}$ K$^{-1}$ and 
$\gamma_0=1.51$ (%for $K_T=268$ GPa and 
taking $C_V=3R$), we obtain a release temperature of $T_2=13200$ K (a reduction of 25\%), in agreement with that calculated using \textsc{hyades}  for this initial condition (Fig. \ref{fig:HYADESambient}).
While this can have a potentially major effect on the starting temperature conditions for finite element models, 
%Cooling by conduction, on much longer timescales than this release, thus begins at a somewhat lower temperature than found in the fixed volume simulations.  However, 
the expansion cooling becomes negligible at lower temperatures, i.e. for Mo at 1000 K the expansion cooling is $<2$ \%. %At higher temperatures, where this effect is particularly pronounced, this model only remains valid so long as no gaps exist in the target which could allow vapourization%.since vapourization could further reduce $V_2$ and hence $T_2$.

%As the interior foil expands outward, the surrounding material is compressed by shock waves moving out from the interior foil, symmetrically (for uniform foil heating) compressing each side of the surrounding medium   These reflect off the higher impedance (diamond) tamper and cause a re-shock state in the medium and a stress maximum in the tamper.  The maximum stress encountered by the tamper occurs at beyond X-X ns).  While the initial pressure in the foil is initially relatively large (X GPa), it is reduced by the hydrodynamic expansion (though not to zero due to the presence of the medium).  The shock transiting the medium and striking the tamper is only X\% of the initial pressure, while the reverberation state approximately twice (?) this.

%FIGURE: Impedance match scenario with and without pre-compression.

As the inertial confinement time in such samples is in the range of picoseconds, radiation pulses significantly longer than the picosecond level will not produce shock waves or large pressure excursions, remaining at or close to the initial pressure.

\section{\label{sec:discussion} DISCUSSION}

%Where heating must be mitigated, the clearest route to reducing this effect, without reducing total power, is increasing the probe beam diameter; similarly, to enhance it, tight focus is required.
%Where heating must be mitigated, the clearest route to reducing this effect, without reducing total power, is increasing the probe beam diameter; similarly, to enhance it, tighter focus may be used; where these 

\subsection{Pulse Train Response}\label{Sec:PulseTrain}

Many high-power x-ray sources involve high-repetition-rate pulse trains, up to the MHz level (pulse separations in the range of hundreds of ns, Table \ref{tab:facilities}), with even faster repetitions possible using, e.g. split delay lines.  %For repeated pulses at this level significant retention of heat between pulses may occur. % accounting for conduction effects that redistribute of heat and a general cooling needs to be accounted for.
For sources operating with high repetition rate, %(~100 ps, Table X) 
faster than the thermal relaxation time of samples (of order 10 $\mu$s in these models), accumulation of thermal energy during a pulse train may occur. 
It may be crucial to consider this energy deposition for serial x-ray measurement (e.g. crystallography\cite{Wiedorn:2018}) applications, even at lower power levels that may normally be considered non-invasive. 
%Given the thermal inertia of the samples modelled here, the repetition rates of many high-power x-ray sources suggest that significant accumulation of heat during a pulse series is possible. %[Give numbers \& Make a Figure for Euro XFEL, consider various response times].
%The timing of the late pulse in temperature could be important as it is timed at roughly at the time further x-ray pulses would strike the target in a bunch mode (about 100 ns, Table X).  
For example, considering the lowest level of irradiation studied here (0.0035 mJ/pulse, Fig. \ref{fig:ResultIntensity}) and assuming a pulse repetition rate of 4.5 MHz (220 ns between pulses, Table \ref{tab:facilities}), the temperature increase between pulses (including heating and cooling) is $\Delta T \simeq 30$ K, implying it would take roughly 50 pulses for an Fe sample to be driven in a step-wise fashion to its melting point (1811 K) from room temperature, in $\sim$11 $\mu$s, assuming the temperature increases linearly with time.
As thermal pressures delivered during pulses have time to dissipate between pulses, concomitant with thermal expansion, this type of heating can be thought of as being nearly isobaric, though the transient thermal pressurization and expansion process itself may have effects on the sample state %including shock heating 
(Sec. \ref{Sec:ResultsHydro}, Sec. \ref{Sec:DiscussionShock}), while residual thermal pressure is possible in well-confined samples\cite{Dewaele:1998}.

\begin{figure}%[htbp]
	\centering
	\includegraphics[width=8.5 cm]{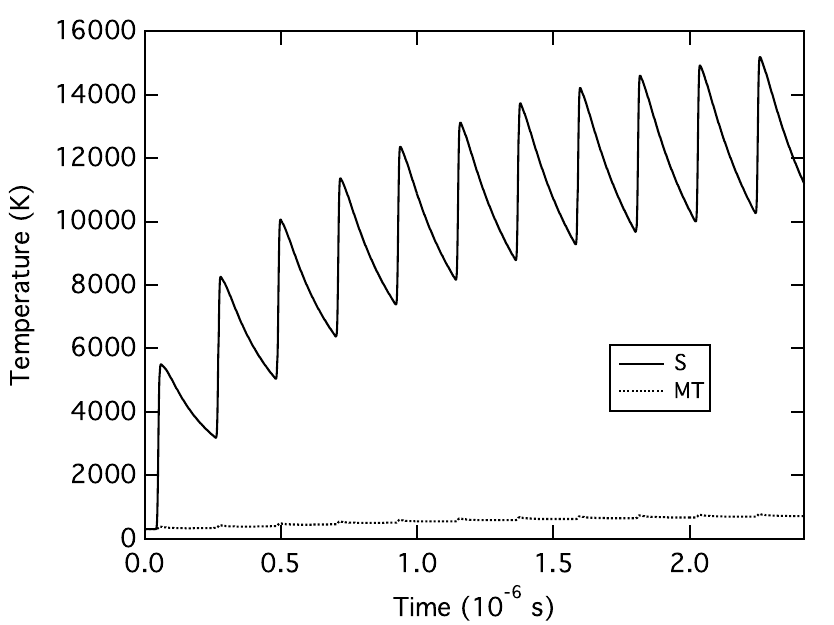} 
	\caption{Stepwise `isobaric' heating by x-ray pulses delivered in a pulse train.  The standard FE model configuration is used, with a 220 ns (4.5 MHz) pulse delay time assumed.  Temperatures at the sample center (S) and medium-tamper interface (MT) are shown, for the first 11 pulses.  Pulse duration is increased to a few ns in this model, to ensure numerical stability in the longer duration simulation.}%, as well as heating before conduction initiation (Fig \ref{fig:ResultsBaseline1}).}
	\label{fig:Stepwise} 
\end{figure}

A representative finite element model of the stepwise heating due to x-ray pulse trains for the baseline experimental arrangement is shown in Fig. \ref{fig:Stepwise}, using serial rather than single exposures at the standard (0.35 mJ/pulse) fluence, assuming a repetition rate of 4.5 MHz. % (220 ns pulse delay).
 The sample temperature grows in a sawtooth fashion, with each pulse producing a new temperature peak followed by a gradual cooling until the next pulse.  Cooling rates increase with temperature, limiting achieved temperatures through a balance between heat added by the x-ray pulses, and energy loss by conduction between pulses, such that peak temperatures rise nonlinearly during the pulse train, and rapidly approach a limiting value.  %Thus, there is likely to be an effective maximum temperature in stepwise heating of this variety.  
In this case the temperature maximum is about three times greater than that achieved following a single pulse.  Similarly, at the lowest fluence (0.0035 mJ/pulse as used in the earlier estimate) the sample would never reach melting, remaining below $\sim$500 K in the limit.

Pulse train experiments may be useful for both probing and heating. For nominally noninvasive probing applications, extending the duration between pulses can reduce the heat accumulated in a fixed target, and ensure the sample temperature rise is minimized at the time of each probing. On longer timescales, the sample temperature at the time of probing is constant, so the data obtained can be treated as isothermal but at an elevated, saturation temperature (after the initial pulses during which stabilization occurs). 
For deliberate heating, minimizing pulse delay can increase the maximum achievable temperature, and the functional length of the pulse train may be the number of pulses required to reach a saturation value (e.g. $\sim$15 pulses for a 4.5 MHz train, Fig. \ref{fig:Stepwise}).
%On longer timescales, the sample temperatures achieved following each pulse would be identical.  
%For 2700 pulses, the European XFEL bunch maximum, peak temperature achieved is $\sim$7 eV over 600 $\mu$s, assuming the sample could survive (e.g. if configured as an anvil cell with eV capability on microsecond timescales \cite{McWilliams:2015}). With higher pulse power, such levels could be reached much faster.  

%FIGURE: STEPWISE 'ISOBARIC' HEATING.
%X curve shows pressure in sample; pressure equilibrium occurs very quickly ($\lessim$ 1 ns) on this timescale.
%Electronic equilibration effects occur on even shorter timescales, i.e. very close to the time of energy delivery ($\lessim$ 1 ps) are not included in T, and also occur instantaneously on this timescale.

%\section{X-ray probing and heating}
%
%A standard type of probing to be performed with strong x-ray beams at a tight focus is x-ray diffraction, e.g. on a polycrystalline sample.  For a given material and x-ray energy, scattering intensity is roughly proportional to $Z^2$ [NeedRefMalcolm?] while absorption is proportional to $Z^4$ (Eq. \ref{Eq:xra}).  Thus, for a given x-ray diffraction (coherent scattering) signal intensity, x-ray heating will be most pronounced for higher Z materials; minimized heating would be obtained for those of low Z.
%
%Naturally, heating in x-ray absorption measurements would be roughly similar for any $Z$, for a given signal intensity.

%Diagnosis of temperature in actual experiments as introduced by x-rays can be accomplished using subsequent beam pulse (such as a two- or more pulse train with delays in the 100 ns range). For example, diffraction can provide an accurate probe of the instantaneous thermal state of the system via detection of thermal expansion [
%http://aip.scitation.org/doi/pdf/10.1063/1.4976541; Dewaele?
%].

\subsection{Target Damage and Mitigation}

Either in a single- or multiple-exposure experiment, the target lifetime can be of central importance.  In a traditional isochoric heating experiment on thin layered targets, the lifetime is set by hydrodynamic expansion of the hot target, occurring as the ions gain energy from electrons and expand into vapor.  By confining the hot target in a tamper, this time can be increased.  Use of very massive tampers surrounding a hotspot can lead to total confinement of even a dense plasma state, and reliable target survival\cite{Vailionis:2011,McWilliams:2015}.  In what follows, basic mechanisms for target failure and their mitigation for long-duration and serial experiments are discussed.
The considerations here apply principally to the effects of a single pulse, inasmuch as the primary damage should occur during the pulse and subsequent thermomechanical relaxation.

\subsubsection{Thermal damage}

Significant damage in targets can result from thermal effects, %e.g. %melting and other phase transformations (e.g. transformation of metastable diamond to graphite).% and sublimation of graphite).
%Primary thermal effects 
which %could 
include reversible and irreversible phase transformation (e.g. melting), reaction, strength reduction (i.e. in the tamper), and for free surfaces, or at gaps, the possibility of vaporization.  While some of these effects are certain to occur in higher-Z (strongly heated) samples, %based on the temperatures achieved, 
the survival of the target assembly will likely depend on tamper integrity.
The temperature at the surfaces of the tamper %, as these will characterize 
generally determine the peak temperatures to which the tampers are subject, and thus %provide an estimate of 
the ability of tampers to survive the thermo-mechanical cycle and successfully confine the sample throughout.
This includes the tamper surfaces facing the sample, heated by close contact with a hot sample layer, and the free surface facing the beam, heated by peak fluence (Figs \ref{fig:ResultsBaseline1} -- %{fig:E1_All} - %\ref{Fig. E2_All}, 
 \ref{fig:HYADESambient}). 
 
Many of the temperature conditions found in these simulations are in principal such that the tampers can survive irradiation.  Except for softer x-rays (Fig. \ref{fig:ResultEnergy}), low-thermal conductivity tampers (Fig. \ref{fig:ResultTamper}) or no interfacial layer (Fig. \ref{fig:ResultThickness}), temperatures remain below probable damage points of the tamper in these experiments even for significant heating in the sample layer (by 10$^3$-10$^4$ K). %Meanwhile high temperatures achieved in the sample can be quenched fairly quickly, preventing serious damage within the sample area itself\cite{McWilliams:2015}
%McWPRL?others.
For the high-thermal conductivity tampers, the tamper temperature remains below graphitization and oxidation points for diamond ($\sim$1000-2000 K), the sublimation point for graphite ($\sim$4000 K) and melting points for Be and Al$_2$O$_3$($\sim$1500-2300 K), for 25 keV radiation (Fig. \ref{fig:ResultTamper}).
For the standard experimental configuration (diamond-alumina-iron and 25 keV x-rays), the tamper begins with only about $\sim$ 2\% of the temperature change in the sample (Fig \ref{fig:ResultIntensity}) and never exceeds this as the target cools.  Even for temperatures exceeding 50,000 K in any sample, diamond tamper temperatures need not exceed 600 -1400 K (Figs \ref{fig:ResultIntensity} and \ref{fig:ResultSample}), low enough to prevent thermal damage, particularly for brief heating. %(diamond melting, graphitization and oxidation occurring at higher temperatures).  For Be and Al$_2$O$_3$ tamper, where melting is a likely survival criterion, melting points (1560 and 2,345 K, respectively), allow up to $\sim$70,000 and $\sim$17,000 K in an Fe sample (Fig \ref{fig:ResultTamper} and Eq. \ref{Eq:IntensityScaling}).
% [wrong?] this temperature in the tamper with a sample temperature of $\sim$2900 K in a Be tamper, for a $\sim$60,000 K Fe sample (Fig \ref{fig:resultstampers} and Eq. \ref{Eq:IntensityScaling})  or $\sim$ 3100
In contrast, the low thermal conductivity plastic tamper (Fig. \ref{fig:ResultTamper}) leads to elevated thermal confinement near the tamper interface with the sample region, and heating of the tamper surface up to $\sim$1200 K for a sample temperature of $\sim$6000 K, well beyond the thermal degradation point of the material ($\sim$670 K for Kapton).
%[http://www.dupont.com/content/dam/dupont/products-and-services/membranes-and-films/polyimde-films/documents/DEC-Kapton-summary-of-properties.pdf
%] .

%Simpler starting materials are likely to undergo more conventional transformations including melting, ablation, and sublimation.

%5 For example, diamond will undergo graphitization (or oxidation) near ambient pressure and melting under pressure\cite{Eggert:2010} at conditions on the order of $\sim$1000 K or more, so ensuring the maximum temperature encountered in diamond tampers does not exceed this limit is crucial for assessing target survival and the level of any damage incurred.  

\begin{figure}%[htbp]
	\centering
	\includegraphics[width=8.5 cm]{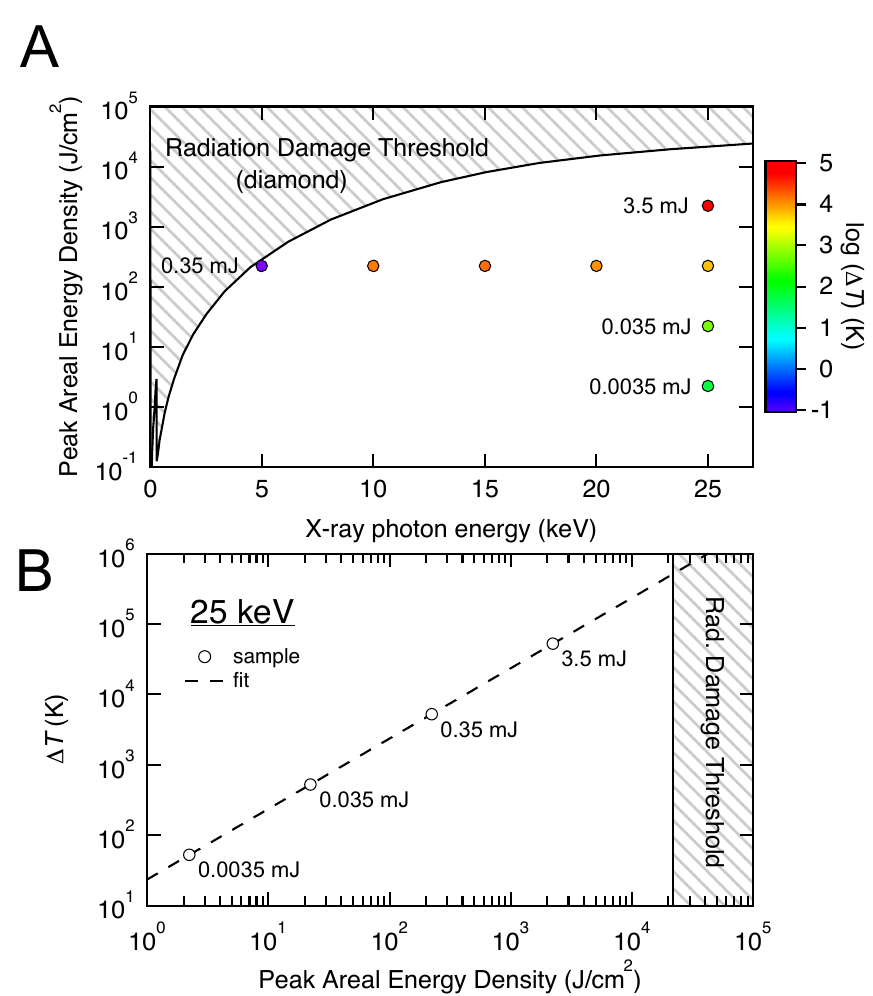} 
	\caption{Comparison of simulated conditions in standard targets (diamond tamper, Al$_2$O$_3$ medium, iron sample) with the `nonthermal' radiative damage threshold predicted for diamond\cite{Medvedev:2013}, given in terms of peak areal energy density $\Lambda_{max}$.  A, Radiation damage threshold of diamond compared with simulated conditions of x-ray energy (Fig. \ref{fig:ResultEnergy}) and fluence (Fig. \ref{fig:ResultIntensity}); color indicates peak temperature achieved in the sample. B, Achieved sample temperature as a function of fluence at 25 keV.  Total energy per pulse is given in mJ.}
	\label{fig:Threshold}
\end{figure}

\begin{figure}%[htbp]
	\centering
	\includegraphics[width=8.5 cm]{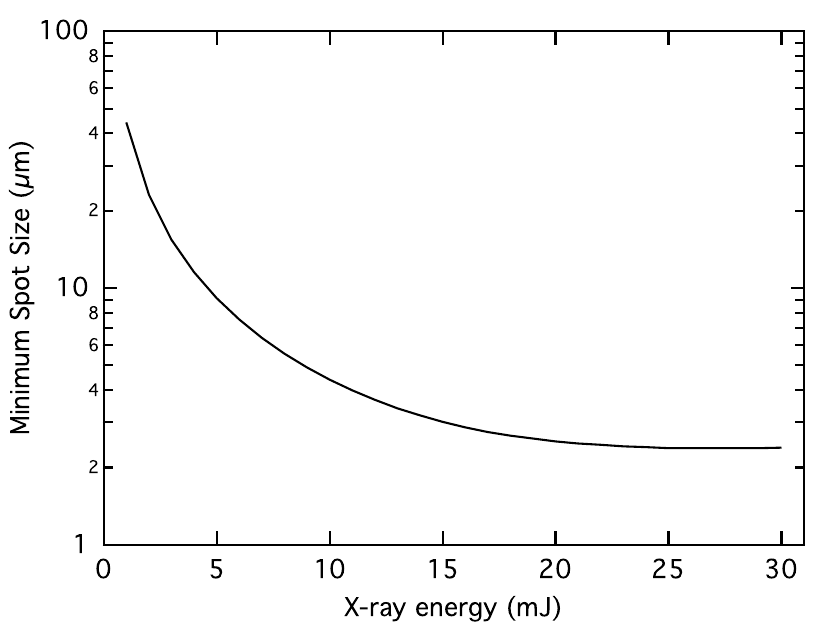} 
	\caption{Effective beam diameter lower limit in diamond assuming a damage threshold of 0.7 eV/atom (graphitization limit\cite{Medvedev:2013,Roth:2018}) and $N=3.5 \times 10^{11}$ photons per pulse (0.06-1.7 mJ/pulse for 1-30 keV%, Eq.\ref{Eq:Nphotons}
).  For the beam diameter used in these simulations, $\sim$12 $\mu$m, the damage threshold is exceeded below 5 keV (see also Fig. \ref{fig:Threshold}) but is within tolerance at higher x-ray energies.  %See also Eqs. \ref{Eq:spotsize}, \ref{eq:PowerEnergy3} and \ref{Eq:Threshold}.
} % \ref{Eq:Nphotons}.}
	\label{fig:SpotSize}
\end{figure}

\subsubsection{Radiation damage}

Ultrahigh intensity laser sources can %dramatically alter the local field in materials and 
have substantial direct influence on materials including radiative damage and electronic excitation:
% incimmediately excite electrons, break bonds and produce instantaneous transformations in materials.
insulators can be rapidly and transiently transformed to metals\cite{Schiffrin:2012}, bonds can be disrupted\cite{Zastrau:2014}, and %ionic 
structural transformations that normally would be sluggish can occur instantaneously\cite{Medvedev:2013}.
Such `non-thermal' radiation effects can be quantified by the amount of energy absorbed per atom, $Q_{atom}$.  From Eq. \ref{eq:attenuation}, integrating over the pulse, and ignoring beam attenuations, the maximum of this quantity is
\begin{equation}
Q_{atom}=\frac{\Lambda_{max} \alpha A}{\rho} \label{Eq:Threshold}
\end{equation}
where $A$ is atomic mass (Eq. \ref{Eq:xra}) and peak energy density per area is $\Lambda_{max}$ (Eq. \ref{eq:PowerEnergy3}).
%for a Gaussian beam spot or, for a square spot of side length $D$, as
%\begin{equation}
%$\Lambda=\frac{E_{pulse}}{D^2}.
%\end{equation}
%A minimum spot size can be thus defined using Eq.\ref{Eq:Nphotons} as
%\begin{equation}
%Q_{atom}=\frac{\alpha N E A}{\rho \pi (D/2)^2}
%\end{equation}
Use of this criterion then leads to rough constraints on acceptable irradiation conditions.

%The absorbed energy $\eta$ can be expressed roughly as (see Eqs. \ref{Eq:Nphotons} and \ref{Eq:xra})
%\begin{equation}
%\eta=\frac{\sigma_{abs} N E}{A}
%\eta=\frac{\alpha N E A}{\rho \pi (D/2)^2}
%\end{equation}
%where $D$ is an effective spot size [\textit{someone check I rewrote this properly}].
%$N$, $E$ and $A$ are the number of photons per pulse, the photon energy and the beam area, respectively, and $\sigma_{abs}$ is photon absorption cross section.   
%where r

Considering again tamper integrity, direct radiative ablation is possible at free surfaces where unconfined atoms may easily escape the target, at $Q_{atom}$ 
$\sim$1 eV;
%, so is most directly relevant to the tamper.  H
however, for the low-Z tampers considered here, such as Be and C polymorphs, this limit is not easily reached\cite{Roth:2018}.
For diamond, nonthermal breakdown of diamond to graphite occurs at relatively lower absorbed energy, $\sim$ 0.7 eV/atom \cite{Medvedev:2013,Roth:2018}.  Even with this more conservative criterion, modeled irradiation conditions remain below the nonthermal damage threshold for diamond\cite{Medvedev:2013} (Fig. \ref{fig:Threshold}A) except possibly at the lowest x-ray energy (5 keV) where, due to considerable direct heating from the x-ray beam (Fig. \ref{fig:ResultEnergy}), the overall damage threshold is likely to be at even lower fluence.  At 25 keV (Fig. \ref{fig:Threshold}B), a diamond tamper could survive irradiation up to iron sample temperatures of $\sim$ 40 eV ($\sim$$5 \times 10^5$ K), and higher-Z sample temperatures in the 100 eV range (c.f. Fig. \ref{fig:ResultSample}); tamper damage risk from heating and shock is likely to be more critical at such conditions.  In summary, direct radiation damage may not be a major factor in target survival and performance.  An effective lower limit on beam diameter to avoid radiation damage in diamond is given in Fig. \ref{fig:SpotSize}.

\subsubsection{Thermo-mechanical damage}\label{Sec:DiscussionShock}

With the rapid, bulk heating of samples occurring faster than pressure wave propagation in our simulations (i.e. $\sigma_t<<d_S/c$), %where $c$ is the pressure wave (sound) speed. Thus 
thermal pressure develops as a consequence of heating.  
The large mechanical stresses associated with target heating can introduce immediate or cumulative damage to targets, including irreversible deformations, flow, fracturing, delamination at interfaces, and spall.  Thus, target survival after a single pulse or series of pulses will depend on the integrity of the target under mechanical stresses as temperature and pressure are raised, and as pressures dissipate hydrodynamically as stress differentials relax (Fig. \ref{fig:HYADESambient}).  The system can exhibit a complex thermomechanical evolution as it moves toward equilibrium if surrounding tampers are sufficiently strong to resist free hydrodynamic expansion.  Mechanical stresses could act in conjunction with direct thermal effects including softening, melting, and vaporization to promote damage.

The magnitudes of mechanical stress initially generated in the target (Eq. \ref{eq:thermalpressure2}) will be similar to those associated with subsequent pressure waves.  In the present examples, while this value can be large, relatively lower stress is applied to the surrounding materials and tampers due to impedance matching requirements.  In our example, for the $\sim$60 GPa initial stress in the Mo sample, %corresponding to a temperature of $\sim$18,000 K 
 shock waves forming in conjunction with the release of the hot sample layer and striking (and reverberating from) the tamper (diamond in this instance) are $\sim$30 GPa in amplitude (Fig. \ref{fig:HYADESambient}). % which is close to (but below) the dynamic yield point of diamond\cite{McWilliams:2010}. 
While tamper temperature is increased somewhat by this shock, in terms of damage threshold it is the pressure perturbation will likely cause the immediate (mechanical) damage.  Notably, the diamond tamper in this case can withstand the shock wave (which falls below the dynamic yielding point\cite{McWilliams:2010}) as well as the subsequent heat wave (Fig. \ref{fig:ResultSample}).  However, shock waves of this amplitude could severely damage other tampers. As the pressure medium controls the shock amplitude, softer media could be used to minimize the shock stress, while complete suppression of shock could be achieved using pulses with durations exceeding the hydrodynamic relaxation times (e.g. synchrotron bunch pulses, Table \ref{tab:facilities}). 

% such as those available in synchrotron bunches.

%It is this stress perturbation (the maximum experienced in the tamper) ; in this example the amplitude of this stress is well below the \cite{McWilliams:2010}. %($\gtrsim$80 GPa for a [100] oriented single crystal) . 
%Such initial stresses on the tamper could thus generally be a main cause of damage; 

%\textbf{DROP: Target interactions leading to tensile stresses may in some cases pose the greatest mechanical risk to the integrity of the targets. For example, tensile stress generated within strongly heated layers as they expand outwards\cite{Ivanov:2003} can potentially tear apart these layers (Fig. \ref{fig:IM}). This can create gaps in otherwise well confined (and even pre-compressed, Sec. \ref{sec:DAC} and Fig. \ref{fig:IM}D) samples.}
%Notably, this can also cause extremely rapid removal of the most severe thermal pressures in the experiment.

%\textbf{FIX CONTACTING LAYER EQ. 1>A AND MAKE REPRESENTATIVE CALC - do this in Results section.}

%\textbf{Fix Mo specific heat and recalculate Mo data; rephrase HYADES part}

%\sub
\subsection{Anvil Cell Configuration}\label{sec:DAC}

As the target configuration discussed here is broadly identical to that of static high pressure cells, this application is considered in detail below.
In an anvil cell type design, the sample is configured to withstand high stresses in the sample area via confinement by thick, hard materials.
Diamond anvils provide unmatched capabilities for pressure application and resistance, for up to $\sim$1000 GPa\cite{Dubrovinsky:2012}, while other strong, low-Z candidates for high-strength tamper-anvils include sapphire (single crystal Al$_2$O$_3$) %corundum) [() []] %(up to 15/20? GPa, [
%Klotz may have record according to Kepa, see mao for this and others)
 and Moissanite (SiC)\cite{Mao:2007}. %[() [
%Chapter 2.09 Theory and Practice ? Diamond-Anvil Cells and Probes for High P?T Mineral Physics Studies, Mao \& Mao 2007 (INCLUDE IN INTRO AS WELL WHEN DAC DISCUSSED), lists all anvils \& max Ps; chris ridely and Konstantin may have a review paper convering this; see also 
%http://www.tandfonline.com/doi/pdf/10.1080/08957959.2015.1009454
%, maybe this covers it, does discuss max P of sapphire of ~15 GPa
  %Following ~\ref{eq:thermalpressure2}, the rough maximum temperature corresponding to these anvils' stress, assuming stress were entirely thermal and isochoric and ignoring thermal softening or melting of anvils, is roughly X K, X K, and X K for diamond, moissanite, and al2O3 for $\alpha=$X, $K_T$=X.  

The prior considerations for limiting target damage suggest that improving sample confinement, i.e. using a pressure cell configuration, could enhance target stability and survival.  
%, is a promising route to achieve improved , %and it is basically identical to the tamped experimental system explored in this study.  
%High confining stress can inhibit cracking or delamination in the hot sample area, preventing vapourisation, and otherwise minimise free expansion of the solid\cite{Dewaele:1998}. 
In this configuration, thermal expansion of the hot sample is limited \cite{Dewaele:1998} ensuring the material remains at or near its initial density regardless of heating.  Cracks or voids which can be present in multilayer target assemblies or ordinarily develop due to thermal stresses can be suppressed. %, and  %preventing vaporization even if extreme temperatures are reached.  
%Anvil cells can sustain significant pressures in samples.  
The ability of anvil cells to resist the heating and associated mechanical stresses in hot samples have long been demonstrated using infrared lasers to heat samples, to temperatures in the range of several eV, over timescales of microseconds and longer\cite{Dewaele:1998,McWilliams:2015,Goncharov:2010}.
With similar conditions of temperature, pressure, and timescale found in the present x-ray heating simulations, many advantages and techniques of the anvil cell configuration may be useful in thickly-tamped target experiments generally.

%to $T=0.1-1$ eV
%The ability of diamond anvil cells to withstand high temperature conditions in samples imposed by laser heating %, including pulsed heating, 
%is also well established, with 
%
%With current , similar to the lifetime of the extreme state in the present x-ray heating approach.
%These advantages, long demonstrated using lasers to introduce thermal stresses in the 0.1-1 eV range to diamond anvil cells\cite{Dewaele:1998,McWilliams:2015,Goncharov:2010}, could thus be instructive.  
%.  

In one possible experiment, a tamped sample could be placed under some small initial stress (to ensure good initial confinement, and void elimination).  Thermal stresses introduced by x-ray heating could be controlled by the anvil's high strength and potential stress resistance.  
%For example, in one potential experiment a tamped sample of an anvil-cell design could be placed under some small initial stress to provide initial confinement; thermal stresses then introduced by heating could then be controlled by the anvil's high strength and peak potential confining stress.
%The anvil cell furthermore allows high initial stress (pre-compression) to be applied prior to the experiment.    
So long as the anvils can withstand the additional mechanical stresses (on the order of GPa or higher for conditions considered here, Sec. \ref{Sec:ResultsHydro}) and any heating (Sec. \ref{Sec:FEResults}), % which is readily achieved) %[, as well as potentially compensate for any tensile stress] (Figs \ref{fig:HYADESambient} and \ref{fig:IM}) 
 the target could be stabilized indefinitely.
%Adding pressure to a diamond-tamper sample can also suppress diamond damage from sample heat by moving heated areas to higher pressures and into the diamond stability field, potentially raising the local damage threshold considerably (e.g. to the melting point under high pressure \cite{Eggert:2010}).
The anvil cell provides a built in way to safely relieve thermal stresses in samples to a mechanical equilibrium confinement state\cite{Dewaele:1998} without hydrodynamic expansion, solving a principal issue in tamped laser-driven targets that may only be partially mitigated by tamping alone.  %[This is discussed in more detail in the following section.]
The extended target stabilization would permit studies over a wide range of timescales, accessing phenomena including electron-ion thermalization, structural transitions and thermal conduction, and enable repeated exposures of the same sample on arbitrary timescales, and sample recovery.
This approach would require some apparatus to apply a compressive force across the target, as in a standard pressure cell configuration, with suitable windows for admittance and observation of radiation. 
%The $\Delta P _V$ assessed in the earlier manner  gives an upper limit on the mechanical stress that might be encountered in such experiments.  Reduction from this value occurs due to the hydrodynamic expansion (nominally to zero stress for free expansion) and decreasing temperatures away from the hotspot need also be considered.  Typical `equilibrium' thermal pressures in a laser-heated diamond anvil cell having similar temperature distributions to those obtained here are $\sim$ 30\% of $\Delta P _V$ \cite{Dewaele:1998}, which accounts for the partial confinement of the sample in the long duration limit after hydrodynamic processes have occurred.  Indeed, 
%Finally/Thus, 

The ability to pre-compress samples to elevated densities can also provide, in conjunction with x-ray irradiation, a route to studying laser-plasma interactions and warm dense matter at conditions of very high density, exceeding that of conventional solid states.  
Static pre-compression of matter to hundreds of GPa confining pressure, or larger using modern double-stage anvils\cite{Dubrovinsky:2012}, is a widely used method, compatible with a variety of strategies to further modulate sample conditions (e.g. temperature) and probe sample properties at extremes. 
Our models demonstrate that coupling a high density sample with intense x-ray irradiation on modern light sources can offer a new approach for exploring ultra-dense and hot states, complementary with dynamic compression and traditional optical-laser-heated DAC, in terms of achievable pressure-temperature-timescale conditions.
Indeed, x-ray heating may serve as an alternative to optical laser-heating\cite{Mao:2007,Dewaele:1998,Goncharov:2010,McWilliams:2015} of anvil cells, 
with the modeled pulsed x-ray heating of samples closely resembling pulsed optical laser heating approaches\cite{Goncharov:2010,McWilliams:2015, Konopkova:2016,Goncharov:2012}, with several key differences.  %exhibits some similarities to laser isochoric heating of foils, though, to date, these 
 Optical heating techniques produce large temperature gradients in samples, i.e. where heat must conduct from a heated surface, and are susceptible to unpredictable coupling related to surface or sample properties; furthermore probes must be carefully aligned with the heated spots.  Hard x-ray heating can in contrast provide homogenous temperatures in the sample bulk on rapid timescales\cite{Nagler:2009}, simple coupling with the sample, and automatic alignment of heating and x-ray probe beams. 
% limit achievable temperature in samples due to screening of laser energy and distribution of energy over larger volumes.
X-ray heating may be particularly useful where introduction of optical laser energy to samples is impractical or impossible, such as where optically opaque anvils are used, e.g. in double-stage anvil\cite{Dubrovinsky:2012} or multi-anvil applications, where the optical damage threshold of anvils may be exceeded in high-energy applications\cite{Armstrong:2010}, or where nominally transparent insulating media transform to opaque conductors during heating\cite{McWilliams:2015}.

Addition of pressure could, at least for the sample interfacing region, serve to elevate the damage thresholds for a diamond tamper, both in terms of its thermal resistance and mechanical resistance. Thermal graphitization is prevented above $\sim$13 GPa where diamond becomes the stable structure of carbon, whereas the melting temperature of diamond at these conditions exceeds 4000 K\cite{Eggert:2010}.  Confining pressure also increases the strength of diamond\cite{McWilliams:2010}, a fact employed in modern anvil cell designs to enhance the potential stress resistance\cite{Dubrovinsky:2012}.

\begin{figure}[htbp!]
	\centering
	\includegraphics[width=8.5 cm]{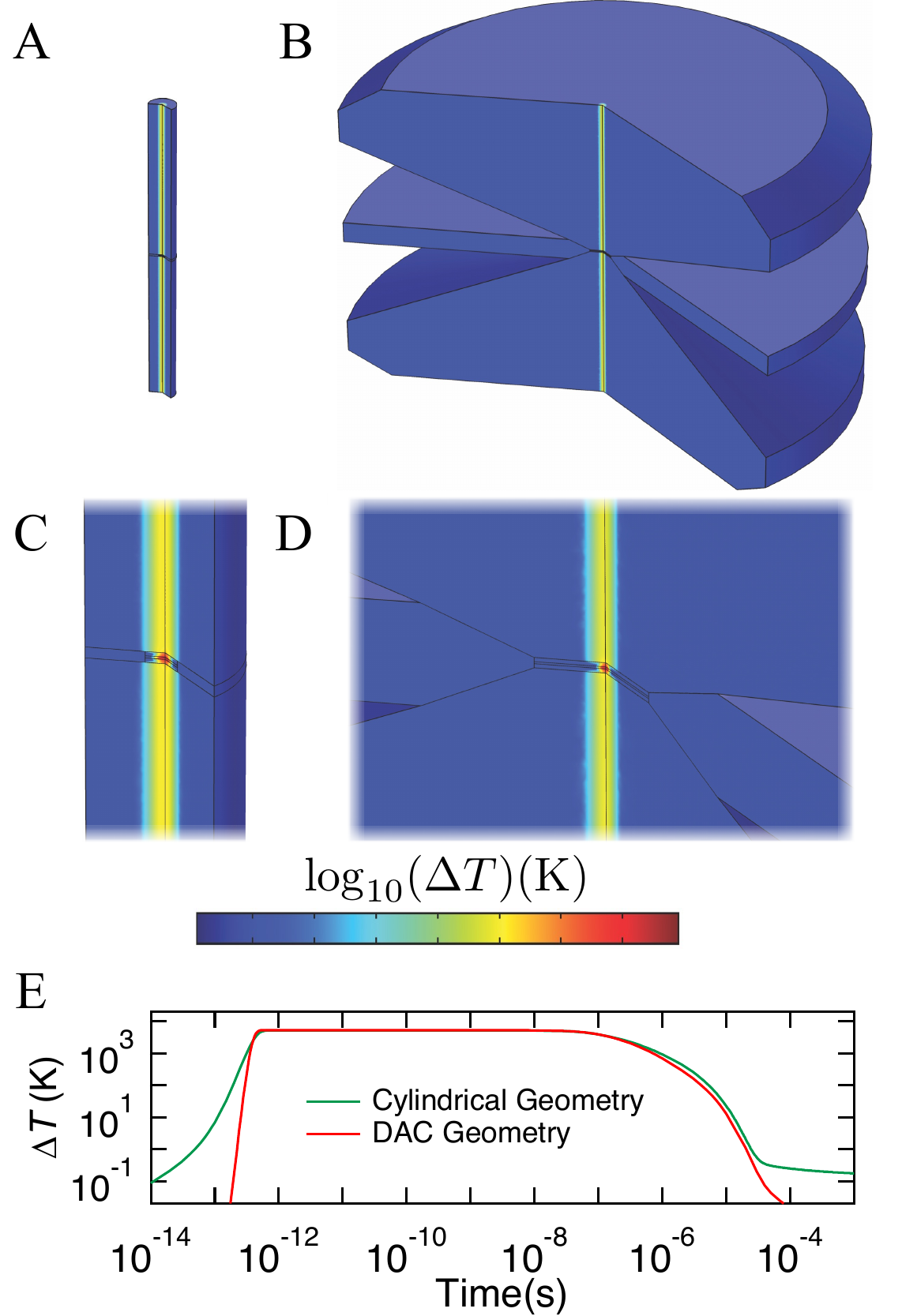} 
	\caption{Comparison between temperature distributions for the main simulation geometry with the standard materials (A) and a comparable simulation for a diamond anvil cell (B). The time of the simulation snapshots in A-D is $1 \times 10^{-7}$ s.  Close-ups of sample regions (C-D) show nearly identical temperature behavior at these early times. A comparison of the temperature history at the sample center shows notable differences in simulated temperature only during heating (a shorter pulse was assumed for the DAC simulation), and very late in the cooling phase.  The latter difference is due to the larger heat sink provided by the full-size target assembly, resulting in lower limiting temperature.  %Slight differences in beam diameter between the simulations are also noted.
	}
	\label{fig:DACcomparison}
\end{figure}

Fig. \ref{fig:DACcomparison} compares two different types of geometry used in our simulations: the first is the cylindrical geometry used in the main simulations, and the second is a representation of an anvil cell.  For similar peak temperatures, there is little difference between the simplified cylindrical model and the more complete model in terms of the temperature evolution of the sample area.  Thus finite-element calculations using the present simple geometry accurately describe the anvil cell design. 
%Comparing the cylindrical symmetry and DAC geometry the changes are in the diffusion part being $1\mu$[s] enough to dissipated all the thermal energy generated to the standard simulation.
%The cylindrical system modelled is found to describe the principal thermal phenomena of the more complex target assembly accurately

\section{\label{sec:conclusion} CONCLUSIONS}

This study describes the thermo-mechanical response of macroscale targets subjected to irradiation by intense, brief x-ray pulses, similar to those now produced by the current generation of x-ray free electron lasers.  These targets use thick, light-element tamper or anvil layers, which are  transparent at hard x-ray energies, to confine a thin target assembly, comprising one or more layers which may be strongly absorbing to x-ray radiation. 
The thermal and mechanical evolution of the x-ray heated target after the rapid deposition of heat is treated using finite-element and radiation-hydrodynamics calculations.  We find that conventional hydrodynamics, classical diffusive heat transfer, and equilibrium thermodynamics can accurately treat the principal thermo-mechanical phenomena for the length and time scales characteristic of such large targets.
%electronic excitation and equilibration processes are assumed to have occurred, such that 
%
%thermodynamic equilibrium.
%

%This approach can allow a sudden creation of high-temperature conditions inside of a bulk target where the extreme state produced is both inertially and thermally confined: the sample is unable to undergo significant expansion from its initial density (i.e. from solid density) due to the large amount of material surrounding it, and hence the experiment is nearly isochoric, while the sample is unable to cool significantly either through expansion or by heat transport processes.

Conditions achieved in the most extreme experiments simulated fall within the regime of warm dense matter, i.e. conditions near or above solid density and temperatures exceeding several eV, where ratios of Coulomb interaction energy to thermal kinetic energy $\Gamma$ (the coupling parameter) and of Fermi energy to thermal energy $\Theta$ (the degeneracy parameter) approach unity.  %[\textit{Sound good?}] 
That these conditions could be sustained for up to microsecond timescales using suitable target configurations offers a potential way to study properties of warm dense matter under total thermodynamic equilibrium conditions, on timescales exceeding those of modern experiments that use laser-driven shock waves or unconfined isochoric heating.  Using thick tampers to apply initial pressure on samples and taking advantage of serial irradiation can enable further exploration of novel regimes of density, temperature, and timescale in warm dense matter.
%Mention somewhere, single-shot experiments can be affected by undesireable heating, e.g. where a probe strikes a dynamically shocked target while it is being continuously observed by other diagnostics, such that the probe influences the other simultaneous probes to the system.
%Also, given that x-ray probes will often be coupled to targets already driven to extreme conditions, the heating behaviour discussed here could occur on a background of major hydrodynamic changes, leading to additional thermal and hydrodynamic perturbations which could produce potentially complex, yet possibly desired experimental conditions.
%
%In serial applications, where exposures may be made on identical or nearby sections of the same target, control of damage may be critical.  
Target survival over one or more exposures is controlled by targets' potential resistance to temperatures on the order of an eV (thousands of degrees Kelvin), mechanical stress close to one million atmospheres (100 GPa), and radiation levels close to damage thresholds, all found to be survivable depending on target design.
%Thermomechanical stresses of order tens of GPa, sufficient to produce high amplitude stress waves including shock waves, can play a significant role in stress and temperature evolution as well as damage.
%Stresses induced by the heating and sudden thermal expansion may be a principal cause of damage in tamped targets, often more so than thermal or radiation effects directly; the mechanical strength of tamper is thus potentially critical. % tampers are low-Z and exhibit modestly high thermal conductivities and hard x-rays are used. 
%The initial stress waves emerging from a confined heated sample would be a probable cause of tamper damage, so 
%, while tension resulting from counter-propogating release waves could 	damage samples.
%, the survival limit is strongly defined by the targets potential resistance to thermal-mechanical stress, shock,  and tension, potentially more than by any direct thermal or radiation damage in many instances considered.  
%Use of thin layers of compressible material surrounding %(or comprising) the 
%heated parts of the target package can dissipate thermal-mechanical stress,  %, is one way to reduce their amplitude.

For thick targets of the considered design ($\mu$m-thick samples with mm-thick tampers), the thermal response due to intense x-ray illumination should be similar at different facilities offering sub-nanosecond pulses, including modern free electron laser and synchrotron sources.  Due to the thermal inertia of samples of this scale, temperatures achieved and cooling behavior are not strongly dependent on pulse lengths, but on total energy dose. Thus modern synchrotron sources with $\sim$100 ps pulse duration may produce a similar level of heating to that at an XFEL with $\sim$100 fs pulses, for equivalent pulse energy.  Heat accumulation over pulse trains with MHz repetition rates characteristic of such facilities can lead to further temperature rise, though this effect is somewhat mitigated by equilibrium between heating and cooling that leads to effectively isothermal experiments on longer timescales.
Thus, consideration of x-ray heating effects may be necessary even in nominally non-invasive x-ray measurements at many modern, high brightness, high repetition-rate x-ray sources, including synchrotron facilities.  
Certain related processes could be more sensitive to the radiation intensity and pulse duration, including shock-wave generation, which would occur under 100-fs XFEL but not 100-ps synchrotron irradiation.%, or initial energy coupling phenomena.  %absorption saturation.

The multilayer target configuration discussed here is informed by, and mimics, the configuration of a static high pressure anvil system, of which the diamond-anvil cell uniaxial press is the most relevant.  
Anvil cells have the ability of preparing initial states of elevated density and pressure in samples, including different structural states, prior to excitation to more extreme states; their wide use in preparing samples for shock-wave compression\cite{Jeanloz:2007, Armstrong:2010}
%[Armstrong, Jeanloz07, Cel10] 
and near-isochoric optical laser heating\cite{Dewaele:1998,Goncharov:2010,McWilliams:2015} experiments suggests many possibilities for accessing otherwise unreachable states of matter with x-ray heating, and for enabling diagnosis of these states by a wide range of radiation types. 
While experience with conventional optical laser heating of anvil cells is relevant, x-ray heating has the potential to bring new advantages for heating pre-compressed matter, including direct volumetric heating, automatic x-ray probe alignment with heated areas, and insensitivity to target optical thresholds. 
The confinement afforded through an anvil cell design is another way to stabilize tamped targets against thermomechanical stress generally and extend experimental lifetimes by limiting them with conductive rather than hydrodynamic dissipation, and ensure target survival for continued exposure and recovery of samples from extremes. % on one part of a sample.  %With timescales of many dynamic heating methods limited by hydrodynamic expansion time, U
 Ultimately experiments must be performed to assess the accuracy of the models developed here, as are currently possible at modern x-ray sources. %  certain key aspects of the physics ( including permanent/persistent/residual changes to samples, nonequilibrium heat transfer, chemistry/phase change, or recrystallization ) but gives a reference condition for the plausible thermal evolution of samples after ultrafast-timescale effects have dissipated.
Further improvements in these models will likely be required to compare with experiments, including coupling of thermomechanical and thermal conductive processes and more accurate treatment of radiation coupling in the sample, %\cite{Vinko:2012,Nagler:2009,Sentoku:2007}, 
which are likely to be essential at higher radiation intensities.

\section*{\label{sec:ack} ACKNOWLEDGEMENTS}

This work was supported by an EPSRC First Grant EP/P024513/1, CONACyT and UAEM{\'e}x, and Leverhulme Trust grant RPG-2017-035.
%Newcastle grant (Natalia), Mexico visitor programme
%, British Council, Carnegie Trust, DESY programs.
Thanks to Y. Ping for providing \textsc{hyades} code, and J. Wark, U. Zastrau, S. Pascarelli, V. Lyamanev, C. Strohm, S. Toleikis, and J. Eggert for helpful discussions.
%Guy from Urbana?, 
%Other coauthors: Zastrau, Eggert?

\bibliography{FELheat_bib}
\bibliographystyle{aipnum4-1}

\end{document}